\documentclass[aps,preprint]{revtex4}
\usepackage[T1]{fontenc}
\usepackage[latin1]{inputenc}
\usepackage{amsmath}
\usepackage{graphics}
\usepackage{graphicx}
\usepackage{amssymb}
\usepackage{dcolumn}
\usepackage{bm}
\usepackage{amsfonts}

\begin{document}

\title{Single production of excited spin-3/2 neutrinos at linear colliders}
\author{O. Çak{\i}r }
\email{ocakir@mail.cern.ch}
\affiliation{Physics Department, Theory Division,  CERN, 1211 Geneva 23, Switzerland}
\affiliation{Ankara University, Faculty of Sciences, Department of Physics, 06100,
Tandogan, Ankara,Turkey}
\author{A. Ozansoy}
\email{aozansoy@mail.cern.ch}
\affiliation{Physics Department, CERN, 1211 Geneva 23, Switzerland}
\affiliation{Ankara University, Faculty of Sciences, Department of Physics, 06100,
Tandogan, Ankara,Turkey}

\begin{abstract}
We study the potential of future high energy $e^{+}e^{-}$ colliders to probe
excited neutrino signals in different channels coming from the
single production process via gauge
interactions. We calculate the production cross
section, decay widths and branching ratios of excited spin-3/2 neutrinos
according to their effective currents and we compare them with that of the spin-1/2 ones.
The signals and corresponding backgrounds are examined in detail to get accessible limits on
the masses and couplings of excited spin-3/2 neutrinos.
\end{abstract}

%%%\date{}
\maketitle

%%%%\pacs{12.60.Rc, 13.85.Rm}

\section{Introduction}

Phenomenologicaly, an excited lepton can be considered as a heavy lepton
sharing leptonic quantum number (flavour) with the corresponding Standard
Model (SM) lepton. Excited leptons ($l^{\ast }$) are suggested by a
composite model of leptons \cite{Terazawa77}. The lepton compositeness would
be the most pronounced in the excess of high-$p_{T}$ events in the
forthcoming high energy experiments. If leptons are composites, they can be
assigned to spin-1/2 bound states, containing three spin-1/2 or spin-1/2 and
spin-0 subparticles. Bound states and/or excited states of spin-3/2 leptons
are possible with three spin-1/2 \cite{Terazawa77} or spin-1/2 and spin-1
subparticles in the framework of compositeness. Composite leptons in the
enlarged groups of standard theory would also imply spin-3/2 leptons \cite%
{Lopes80, Tosa85}. Further motivation for spin-3/2 particles comes from the
supergravity (SUGRA) where spin-3/2 gravitino is the superpartner of the
graviton \cite{Freedman76}. Massive spin-3/2 excited neutrinos with the
analogous heavy spin-1/2 excited ones can be produced at future high energy
colliders through their effective interactions with the ordinary leptons. In
the electron-positron collisions, pair production of excited spin-3/2
neutrinos can also be performed, but it is limited kinematically by $m^{\ast
}<\sqrt{s}/2$, while single production directly can reach masses as high as $%
\sqrt{s}$. Furthermore, excited neutrino could contribute to the pair
production of charged weak bosons through its exchange in the $t-$channel
\cite{walsh03}.

The mass limits for excited spin-1/2 neutrino from its single production
search are $m^{\ast }>190$ GeV and $m^{\ast }>102.6$ GeV from their pair
production \cite{Achard03} assuming $f=-f^{^{\prime }}$.
Relatively small
mass limits are obtained for $f=f^{^{\prime }}$, where $f$ and $f^{^{\prime
}}$ are the scaling factors for the gauge couplings of $SU(2)$ and $U(1)$.
In this case, the mass limits for the excited spin-1/2 neutrinos are:
$m_{\nu_e^*}>101.7$ GeV, $m_{\nu_\mu^*}>101.8$ GeV and $m_{\nu_\tau^*}>92.9$
GeV \cite{Achard03}.
Recently, a search for the excited spin-1/2 neutrinos has been performed by
H1 Colloboration \cite{Aaron08} assuming $f=f^{^{\prime }}$ and $f/\Lambda
=1/m^{\ast }$, with an exclusion limit for the mass range of excited
neutrino $m^{\ast }<213$ GeV at $95\%$ C.L.
Excited spin-1/2 leptons was
studied at hadron colliders in \cite{Baur90} and \cite{Eboli02, Cakir03}
by taking into account possible backgrounds. These studies have shown that
excited spin-1/2 lepton masses up to $2$ TeV can be probed at the LHC.
An analysis of the production
and decay processes of single heavy spin-3/2 neutrinos was performed in
\cite{Choudhury85, Almeida96} in the frame of two phenomenological currents
without taking into account possible backgrounds from the collisions.

In this study, we consider excited neutrino production in more detail (since
excited spin-3/2 neutrinos are least studied compared to the excited
spin-1/2 ones).
We take into account the signals in different decay channels of
excited neutrinos as well as the corresponding backgrounds at the
International Linear Collider (ILC) \cite{Loew03} with $\sqrt{s}=0.5$ TeV
and Compact Linear Collider (CLIC) \cite{Assmann00} with an optimal design
energy of $\sqrt{s}=3$ TeV. We present the missing transverse
momentum distributions for single production of excited spin-3/2 and
spin-1/2 neutrinos. The shape of these distributions
would help to discriminate the signals from the background.

\section{Phenomenological Currents}

\begin{figure}[ptbh]
\includegraphics[width=8cm,height=6cm]{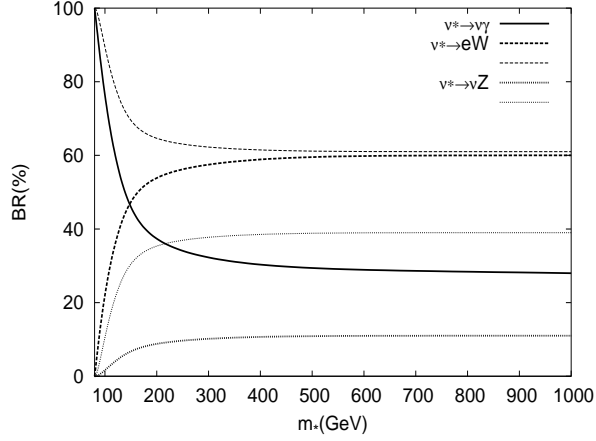}
\caption{The branching ratios ($\%$) depending on the mass of excited
spin-1/2 neutrino for $f=f^{\prime}=1$ (thin lines) and $f=-f^{\prime}=1$%
(thick lines).}
\label{fig1}
\end{figure}

First, we consider the interaction between a spin-1/2 excited neutrino,
gauge boson ($V=\gamma,Z,W^{\pm}$) and the SM lepton described by the
effective current:
\begin{equation}
J_{1/2}^{\mu}=\frac{g_{e}}{2\Lambda}\overline{u}(k,1/2)i\sigma^{\mu\nu}q_{%
\nu }(1-\gamma_{5})f_{_{V}}u(p,1/2)  \label{1}
\end{equation}
where $\Lambda$ is the scale of new physics responsible for the new
interactions, $k,p$ and $q$ are the four-momentum of SM neutrino (electron),
excited spin-1/2 neutrino and gauge boson $\gamma,Z$ ($W^{\pm}$),
respectively. $g_{e}$ is the electromagnetic coupling
constant, $g_{e}=\sqrt{4\pi\alpha}$.
$f_{V}$ is the electroweak coupling parameter corresponding to
the vector boson $V$. In
Eq. (1), $%
\sigma^{\mu\nu}=i(\gamma^{\mu}\gamma^{\nu}-\gamma^{\nu}\gamma^{\mu})/2$
with $\gamma^{\mu}$ being the Dirac matrices. For an excited spin-1/2 neutrino,
three decay channels are
possible: radiative decay $\nu^{\ast}\rightarrow\nu\gamma$, neutral weak
decay $\nu^{\ast}\rightarrow\nu Z$, charged weak decay $\nu^{\ast}%
\rightarrow eW.$ Neglecting the ordinary lepton masses we find decay widths
as
\begin{equation}
\Gamma(l^{\ast(1/2)}\longrightarrow lV)=\frac{\alpha m^{\ast3}}{4\Lambda^{2}
}f_{V}^{2}(1-\frac{m_{V}^{2}}{m^{\ast2}})^{2}(1+\frac{m_{V}^{2}}{2m^{\ast2}})
\end{equation}
where $f_{\gamma}=(f-f^{\prime})/2$ ; $f_{Z}=(f\cot\theta_{W}+f^{\prime}
\tan\theta_{W})/2$ ; $f_{W}=f/\sqrt{2}\sin\theta_{W}$; $m_{V}$ is the mass
of the gauge boson. The parameters $f$ and $f^{\prime}$ are determined by
the composite dynamics, and they can be defined as $q^{2}-$ dependent form
factors. In the literature, they are often taken as $f=f^{\prime}=1$ or $%
f=-f^{\prime}=1$ with $\Lambda=m^{\ast}$. The total decay width of the
excited spin-1/2 neutrino is $\Gamma=3.4(6.9)$ GeV for $m^{\ast}=0.5(1)$ TeV
at $f=f^{\prime}=1$ and $\Lambda=m^{\ast}$. The decay widths for
excited spin-1/2 neutrino for the mass range
$\ 0.2-3$ TeV are given in Table \ref{table1}.
The branching ratios of the excited spin-1/2 neutrinos into
SM leptons and gauge bosons are given in Fig. \ref{fig1}. One may note that
the electromagnetic interaction between excited neutrino and ordinary
neutrino vanishes for $f=f^{\prime}$. As seen from Fig. \ref{fig1}, for $%
f=f^{\prime }=1$ charged current decays become dominant for higher mass
values $m^{\ast }>150$ GeV, while $f=-f^{\prime}=1$ the branching ratio is $%
\sim60\%$ for $eW$ channel. Their relative importance depends on the gauge
boson mass and couplings. At $f=-f^{\prime}=1$ the branchings will be $60\%$
for $W$ channel, $12\%$ for $Z$ channel and $28\%$ for $\gamma$ channel at
higher excited spin-1/2 neutrino masses ($m^{\ast}>500$ GeV). Therefore the
signature of $\nu^{\ast}\rightarrow eW$ is preferable in both cases for the
investigation of excited neutrino in future linear collider experiments.

Second, the phenomenological currents for the interactions among spin-3/2
excited neutrino, gauge boson and SM lepton are given by
\begin{equation}
J_{1}^{\mu}=g_{e}u(k,1/2)(c_{1V}-c_{1A}\gamma_{5})u^{\mu}(p,3/2)
\end{equation}
\begin{equation}
J_{2}^{\mu}=\frac{g_{e}}{\Lambda}u(k,1/2)q_{\lambda}\gamma^{\mu}(c_{2V}
-c_{2A}\gamma_{5})u^{\lambda}(p,3/2)
\end{equation}
\begin{equation}
J_{3}^{\mu}=\frac{g_{e}}{\Lambda^{2}}u(k,1/2)q_{\lambda}i\sigma^{\mu\nu}
q_{\nu}(c_{3V}-c_{3A}\gamma_{5})u^{\lambda}(p,3/2)
\end{equation}
where $u^{\mu}(p,3/2)$ represents the Rarita-Schwinger vector-spinor \cite%
{Schwinger41}, $c_{iV}$ and $c_{iA}$ are the vector and axial vector
couplings, the four momenta $q_{\alpha}=(p-k)_{\alpha}$ belongs to the vector boson.
A spin-3/2 excited
neutrino ($\nu^{\ast}$) decays via two-body process according to the
phenomenological currents Eq.(3-5). The radiative decay widths of excited spin-3/2
neutrinos for three different currents are given as
\begin{align}
\Gamma_{1}(\nu^{\ast(3/2)} & \longrightarrow\nu\gamma)=\frac{\alpha}{4}%
(c_{1V}^{\gamma^{2}}+c_{1A}^{\gamma^{2}})m^{\ast} \\
\Gamma_{2}(\nu^{\ast(3/2)} & \longrightarrow\nu\gamma)=\frac{\alpha}{24}%
(c_{2V}^{\gamma^{2}}+c_{2A}^{\gamma^{2}})m^{\ast}(\frac{m^{\ast}}{\Lambda }%
)^{2} \\
\Gamma_{3}(\nu^{\ast(3/2)} & \longrightarrow\nu\gamma)=\frac{\alpha}{48}%
(c_{3V}^{\gamma^{2}}+c_{3A}^{\gamma^{2}})m^{\ast}(\frac{m^{\ast}}{\Lambda }%
)^{4}
\end{align}
and the weak decay widths of excited spin-3/2 neutrinos are given by
\begin{align}
\Gamma_{1}(\nu^{\ast(3/2)} & \longrightarrow lV)=\frac{\alpha}{48}%
(c_{1V}^{2}+c_{1A}^{2})m^{\ast}\frac{(1-\kappa)^{2}}{\kappa}(1+10\kappa
+\kappa^{2}) \\
\Gamma_{2}(\nu^{\ast(3/2)} & \longrightarrow lV)=\frac{\alpha}{48}%
(c_{2V}^{2}+c_{2A}^{2})m^{\ast}(\frac{m^{\ast}}{\Lambda})^{2}\frac {%
(1-\kappa)^{4}}{\kappa}(1+2\kappa) \\
\Gamma_{3}(\nu^{\ast(3/2)} & \longrightarrow lV)=\frac{\alpha}{48}%
(c_{3V}^{2}+c_{3A}^{2})m^{\ast}(\frac{m^{\ast}}{\Lambda})^{4}(1-\kappa
)^{4}(2+\kappa)
\end{align}
where $\kappa=(m_{V}/m^{\ast})^{2}$ and $m_{V\text{ }}$ is the mass of the
vector boson ($W^{\pm}$ or $Z$).

\begin{figure}[ptbh]
\begin{center}
\includegraphics[
height=6cm,
width=10cm
]{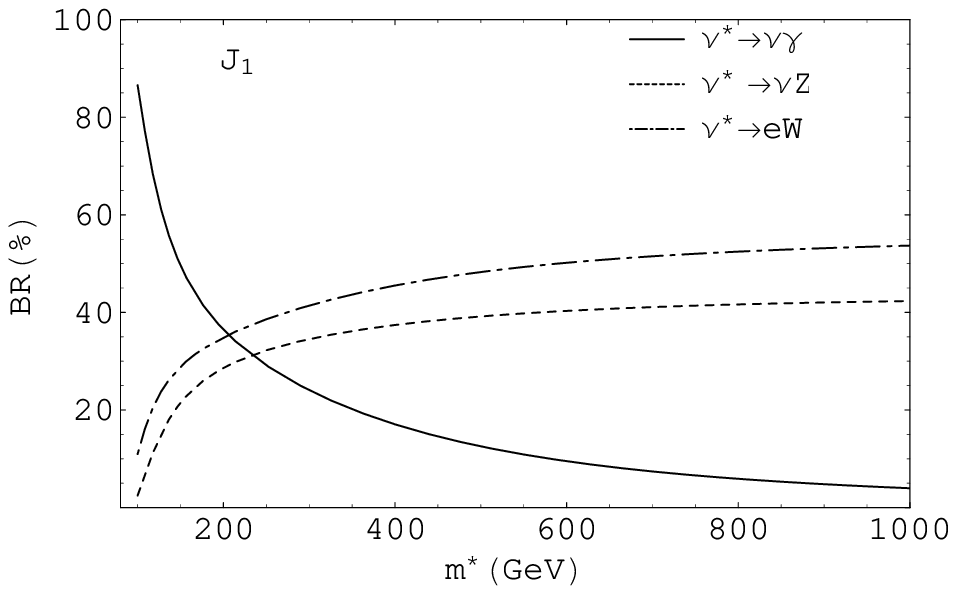}
\par
\includegraphics[
height=6cm,
width=10cm
]{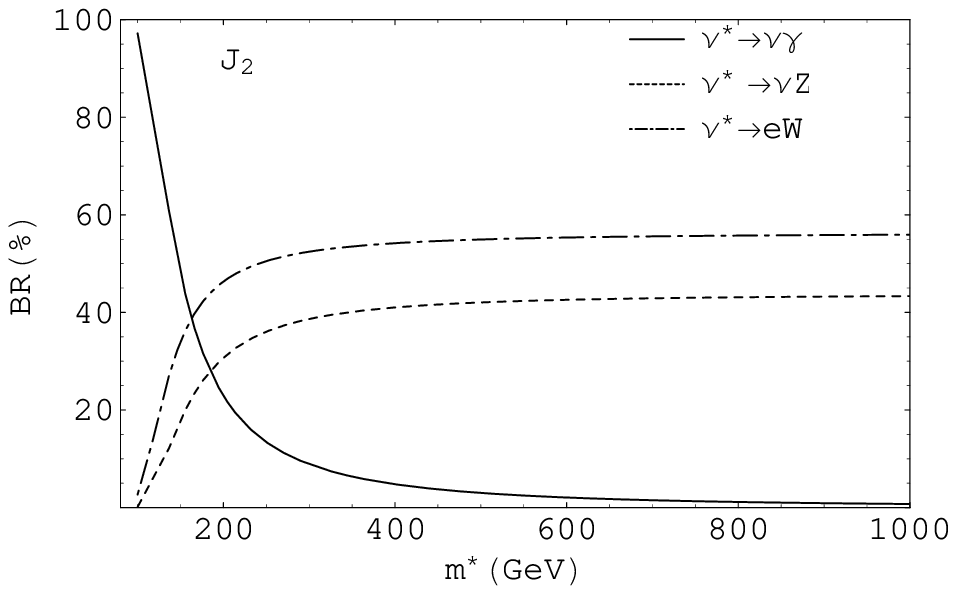}
\par
\includegraphics[
height=6cm,
width=10cm
]{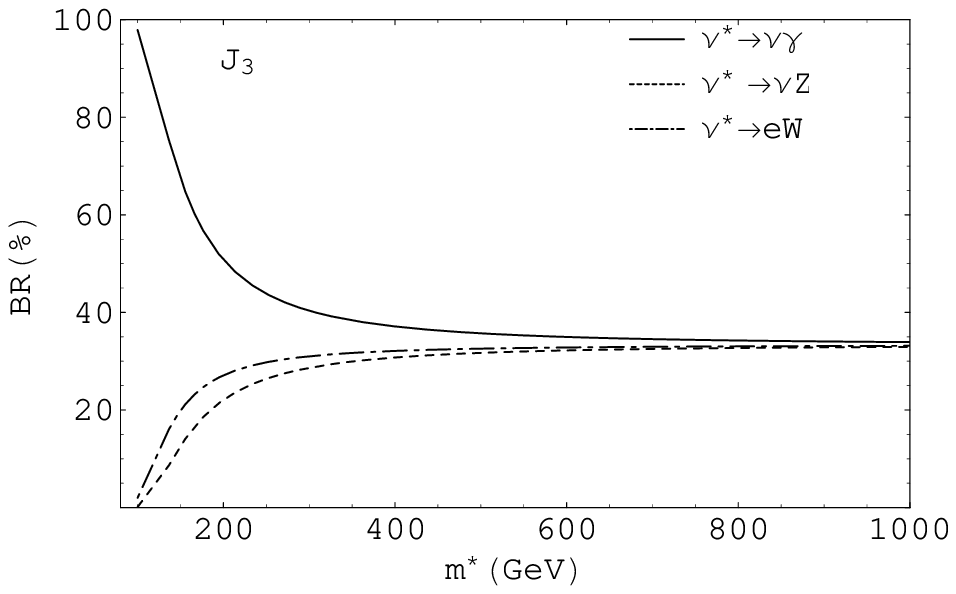}
\end{center}
\caption{The branching ratios ($\%$) of excited spin-3/2 neutrinos depending on their mass
values. The plots (up to down) correspond to the currents $J_1$, $J_2$ and $J_3$ with couplings
$c_{iV}=c_{iA}$ and $\Lambda=m^\ast$.}
\label{fig2}
\end{figure}

The relative importance of higher dimension operators is not
essential when $\Lambda>m^{\ast}.$ Taking $\Lambda=m^{\ast}=0.5(1)$ TeV and $%
c_{iV}=c_{iA}=0.5$ \ we find the total decay widths of the excited spin-3/2
neutrinos as $\Gamma_{1}=3.9(24.6)$ GeV, $\Gamma_{2}=2.7(22.2)$ GeV and $%
\Gamma_{3}=0.23(0.48)$ GeV for the currents $J_{1}$, $J_{2}$ and $J_{3}$,
respectively. The decay widths of excited spin-3/2 neutrino with $%
c_{iV}=c_{iA}=0.5$ for different mass values are shown in Table \ref{table1}.
One can notice that even at $\Lambda=m^{\ast\text{ }}$ and equal
couplings, the values of the decay widths become different. Because, there
are different radiative contributions for each currents and the $\kappa$
term effects differently these channels. The corresponding branchings are
given in Fig. \ref{fig2}. For equal couplings $c_{iV}=c_{iA}=0.5$
and $\Lambda=m^{\ast}$ the
branchings for the weak decays corresponding to the current $J_{1}$ and $%
J_{2}$ becomes dominant for $m^{\ast}\gtrsim 200$ GeV. The radiative and the
weak decay channels with the same couplings have equal probability for the
current $J_{3}$ when $m^{\ast}\gtrsim 500$ GeV.

\begin{table}[tbp]
\caption{Excited spin-3/2 neutrino decay widths depending on their masses for $c_{iV}=c_{iA}=0.5$. The numbers in
the second column show the results for excited spin-1/2 neutrinos with
$f=-f^{\prime }=1$ ($f=f^{\prime }=1$)
at $\Lambda =m^{\ast }$. }
\label{table1}
\begin{tabular}[t]{|c|c|c|c|c|}
\hline
$m^{\ast }$ (TeV) & $\Gamma _{J(1/2)}$ (GeV) & $\Gamma _{J_{1}(3/2)}$ (GeV)
& $\Gamma _{J_{2}(3/2)}$ (GeV) & $\Gamma _{J_{3}(3/2)}$ (GeV) \\ \hline
$0.2$ & $1.15$ $(1.03)$ & $0.54$ & $0.14$ & $0.06$ \\ \hline
0.3 & $1.93$ $(1.85)$ & $1.22$ & $0.55$ & $0.12$ \\ \hline
0.4 & $2.67$ $(2.61)$ & $2.29$ & $1.36$ & $0.18$ \\ \hline
0.5 & $3.39$ $(3.35)$ & $3.89$ & $2.71$ & $0.23$ \\ \hline
0.75 & $5.18$ $(5.15)$ & $11.12$ & $9.31$ & $0.36$ \\ \hline
1.0 & $6.95$ $(6.93)$ & $24.62$ & $22.20$ & $0.48$ \\ \hline
1.5 & $10.43$ $(10.45)$ & $78.89$ & $75.24$ & $0.78$ \\ \hline
2.0 & $13.97$ $(13.98)$ & $183.48$ & $178.61$ & $0.97$ \\ \hline
2.5 & $17.49(17.47)$ & $355.16$ & $349.07$ & $1.22$ \\ \hline
3.0 & $20.99$ $(20.98)$ & $650.72$ & $603.41$ & $1.46$ \\ \hline
\end{tabular}
\end{table}

\section{Cross Sections}

\begin{figure}[tbph]
{}\includegraphics[
height=6cm,
width=8cm
]{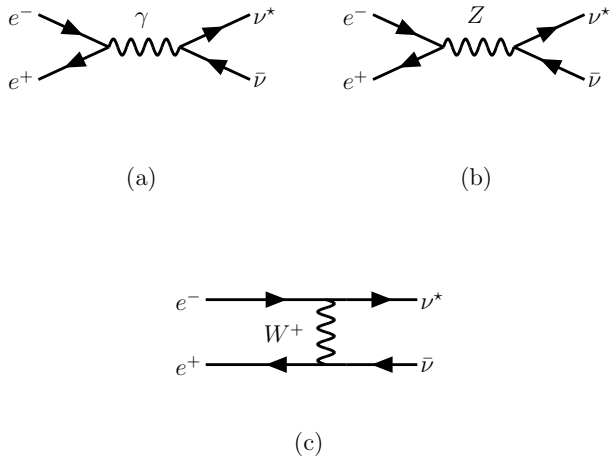}
\caption{Feynman diagrams for single excited spin-3/2 or spin-1/2 neutrino
production in $e^{+}e^{-}$ collisions.}
\label{fig3}
\end{figure}

The spin-1/2 and spin-3/2 excited neutrinos can be produced singly at future
$e^{+}e^{-}$ colliders, namely ILC and CLIC. The Feynman diagrams
contributing to the single production of excited neutrinos via the $s$%
-channel $Z,\gamma$ exchange and $t$-channel $W$ exchange are shown in
Fig. \ref{fig3}.

One may note that pair production of excited neutrino in the $t$-channel $W$
exchange is also possible with the currents given in the previous section.
In the $s$-channel pair production of excited neutrinos of any type could
also be possible if their interactions with photon and $Z$ boson are
defined. On the other hand, the single production of $\nu _{\mu }^{\ast}$
and $\nu _{\tau }^{\ast }$ with the $s$-channel process is possible
with and without the lepton
flavour violating coupling (LFV).

Recently, spin-1/2 excited neutrino have been studied at the ILC energy $%
\sqrt{s}=0.5$ TeV in \cite{Cakir04}. The $\nu^{\ast}\rightarrow eW$ signal
has been searched for an easy identification and accurate measurements in a
linear collider environment. The results show that spin-1/2 excited neutrino
can be probed up to the mass $m^{\ast}=450$ GeV in the charged weak decay
channel when the coupling parameters are taken as $f=f^{\prime}=0.1$ for $
\Lambda=m^{\ast}$.

\begin{figure}[tbph]
\includegraphics[
height=6cm,
width=8cm
]{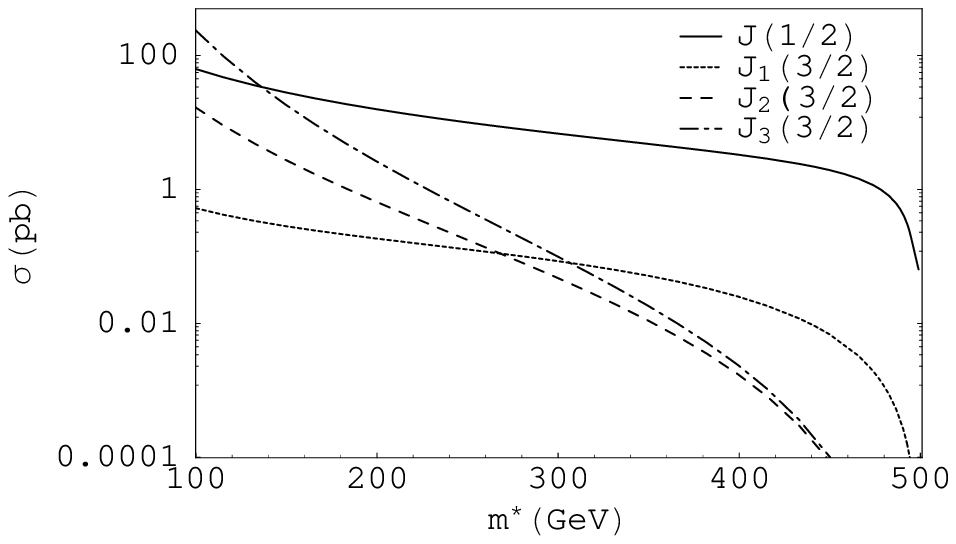}\includegraphics[
height=6cm,
width=8cm
]{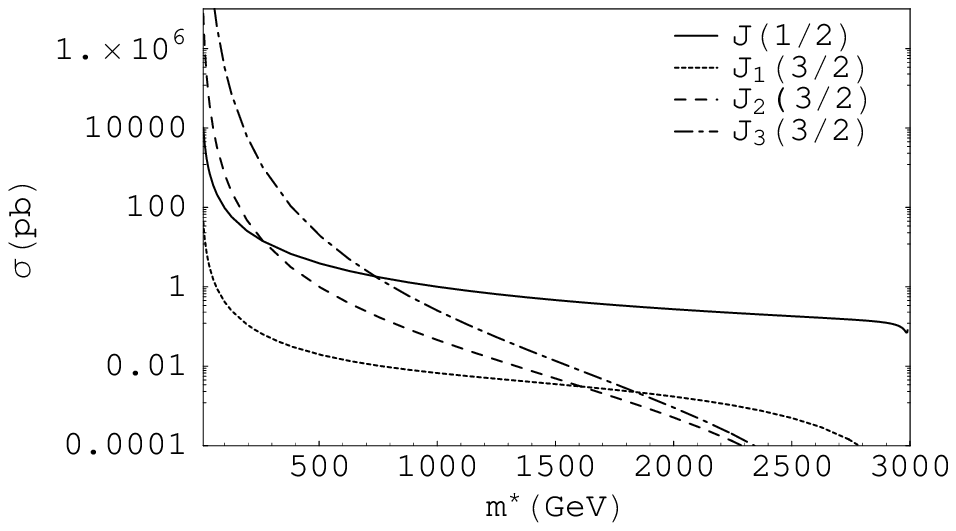}
\caption{Excited neutrino production cross section as a function of the mass
at the ILC\ energy $\protect\sqrt{s}=0.5$ TeV (left panel) and CLIC energy $%
\protect\sqrt{s}=3$ TeV (right panel). Solid, dotted, dashed and dot-dashed
lines denote spin-1/2 current J(1/2) for $f=-f^{\prime }=1$
and the spin-3/2 with $J_{1},$ $J_{2}$, $J_{3}$ currents
for $c_{iV}^{\gamma }=c_{iA}^{\gamma}=0.5$, respectively. }
\label{fig4}
\end{figure}

\begin{figure}[tbph]
\includegraphics[
height=6cm,
width=8cm
]{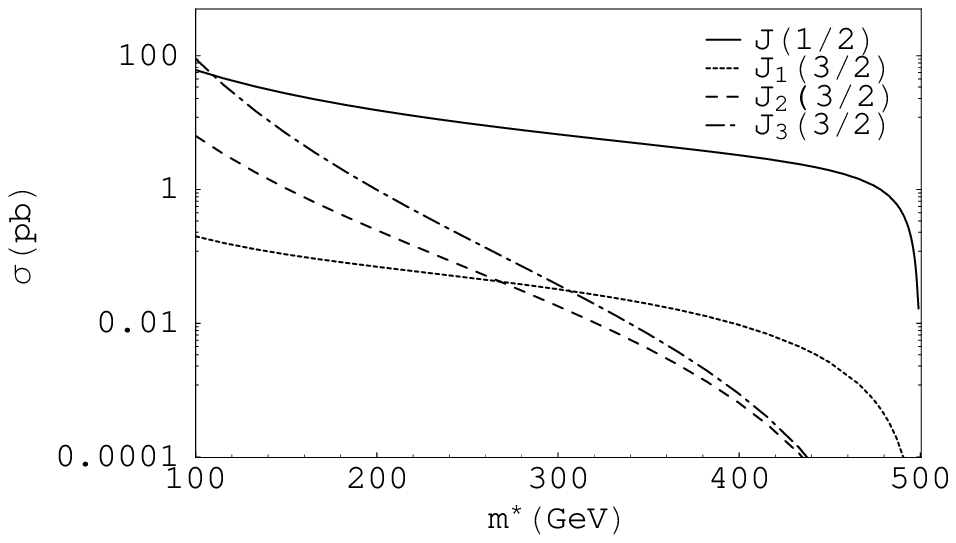}\includegraphics[
height=6cm,
width=8cm
]{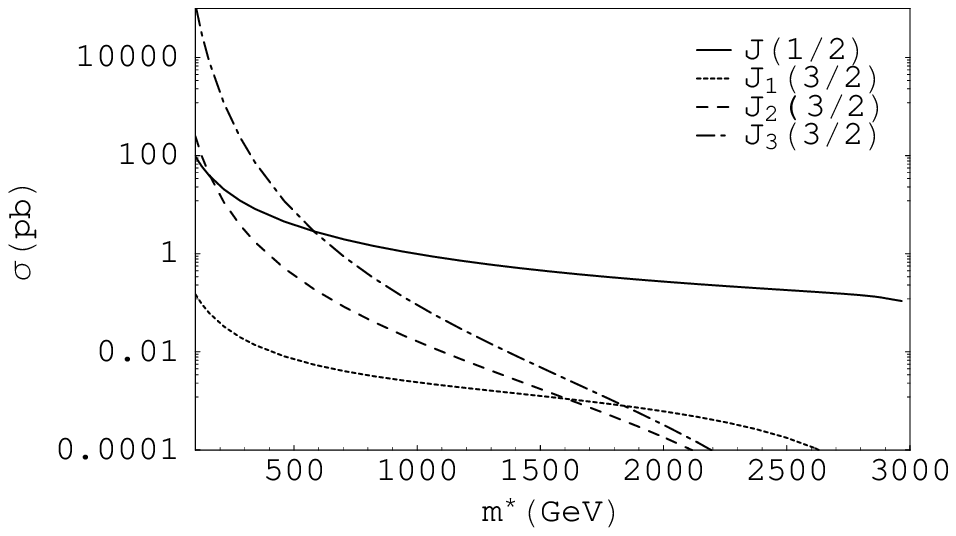}
\caption{Excited neutrino production cross section as a function of the mass
at ILC with $\sqrt{s}=0.5$ TeV (left panel) and CLIC with
$\sqrt{s}=3$ TeV (right panel). Solid, dotted, dashed and dot-dashed
lines denote spin-1/2 and spin-3/2 currents $J(1/2)$, $J_{1},$ $J_{2}$, $J_{3}$
currents, respectively. Here, we take $f=f'=1$ and $c_{iV}^Z=c_{iA}^Z=0.5$ with
$\Lambda =m^{\ast }.$}
\label{fig5}
\end{figure}

\begin{figure}[tbph]
\includegraphics[
height=6cm,
width=8cm
]{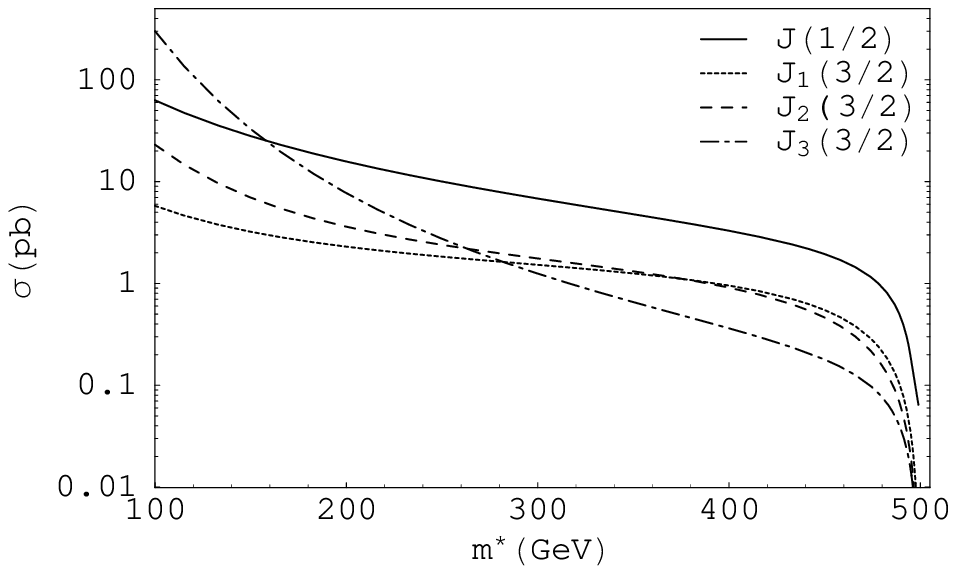}\includegraphics[
height=6cm,
width=8cm
]{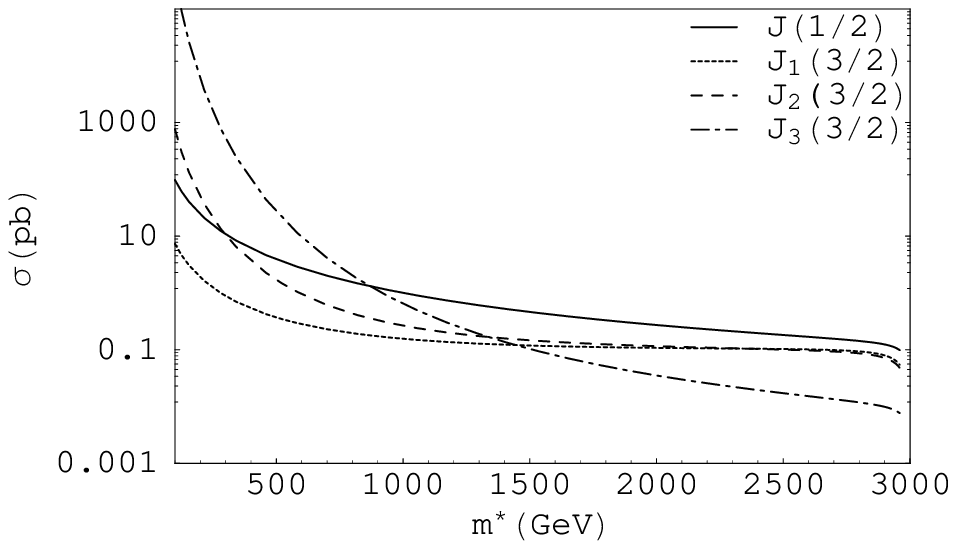}
\caption{Excited neutrino production cross section as a function of the mass
at ILC (left panel) and CLIC (right panel). Solid, dotted, dashed and dot-dashed
lines denote spin-1/2 and spin-3/2 currents
for $f=-f^{\prime }=1$ and $c_{iV}^{W}=c_{iA}^{W}=0.5$, respectively. Here we
take $\Lambda =m^{\ast }$. }
\label{fig6}
\end{figure}

In order to calculate the cross sections for $e^{+}e^{-}\rightarrow
\nu_{e}^{\ast}\bar{\nu}_{e}$ processes for three phenomenological spin-3/2 currents $
J_{1},J_{2}$ and $J_{3}$, we integrate over the Mandelstam variable $t$ (or $
\cos\theta$) of the diffential cross section
\begin{equation}
\frac{d\sigma }{dt}^{(k)}=\frac{{g}_{{e}}^{2\,}}{24{m}^{\ast 2}\,\pi \,s^{2}}
\sum_{i\leq j=1,3}
\frac{T_{ij}^{(k)}}{P_{ij}^{(k)}}
\end{equation}
where we use the expressions $T_{ij}^{(k)}$ and $P_{ij}^{(k)}$ as given in
the Appendix. Total cross sections as a function of excited neutrino mass
are presented in Figs. \ref{fig4}-\ref{fig6} at center of mass energy$\sqrt{s}=0.5$
and $\sqrt{s}=3$ TeV.

One can see from Figs. \ref{fig4} and \ref{fig5}, the excited spin-3/2 neutrinos with
current $J_{3}$ (having only $c_i^\gamma$ or $c_i^Z$ couplings) has a cross
section larger than the other two currents ($J_1$ and $J_2$) when its mass
below $1.8$ TeV and when they are produced at $\sqrt{s}=3$ TeV. The cross section for the $%
\nu^\ast(J_3)$ via $t-$channel $W-$exchange becomes more visible below $%
m^\ast<1.4$ TeV, as shown in Fig. \ref{fig6}. Depending on the couplings $%
c_{iV},c_{iA}$ their (currents) relative importance
(changes in the cross section for the interested mass range)
become more pronounced compared to
the excited spin-1/2 neutrinos.

\begin{figure}[tbph]
\includegraphics[
height=6cm,
width=8cm
]{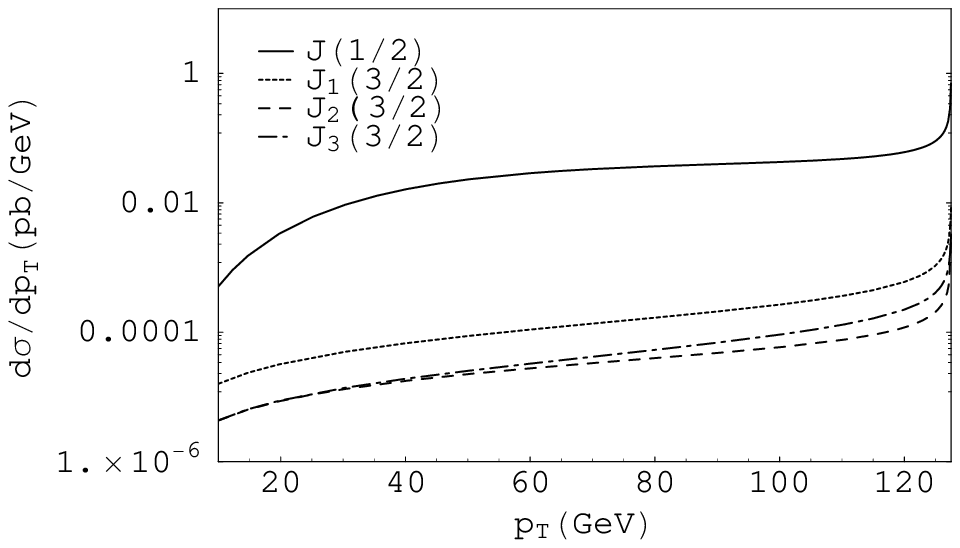}\includegraphics[
height=6cm,
width=8cm
]{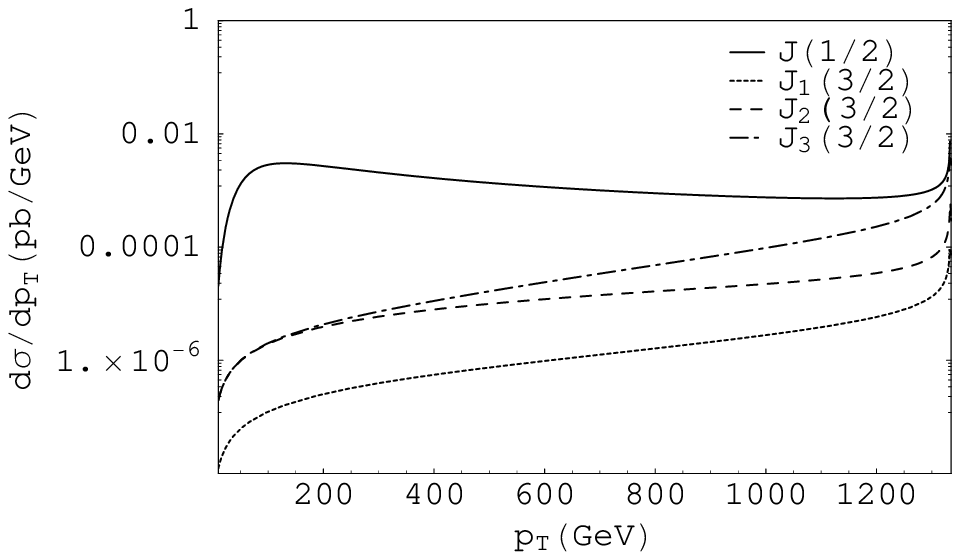}
\caption{The missing $p_{T}$ distribution taking into account
the coupling $c_{iV}^\gamma=c_{iA}^\gamma=0.5$ for spin-3/2 currents and $f=-f'=1$ for
spin-1/2 current with $\Lambda=m^*$ where $m^*=350$ GeV for ILC and
$m^*=1$ TeV for CLIC.}
\label{fig7}
\end{figure}

\begin{figure}[tbph]
\includegraphics[
height=6cm,
width=8cm
]{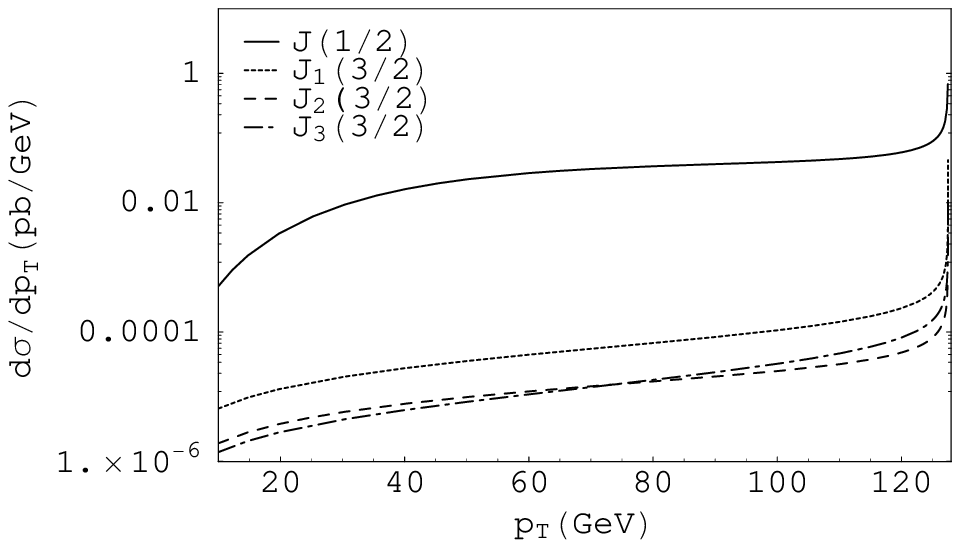}\includegraphics[
height=6cm,
width=8cm
]{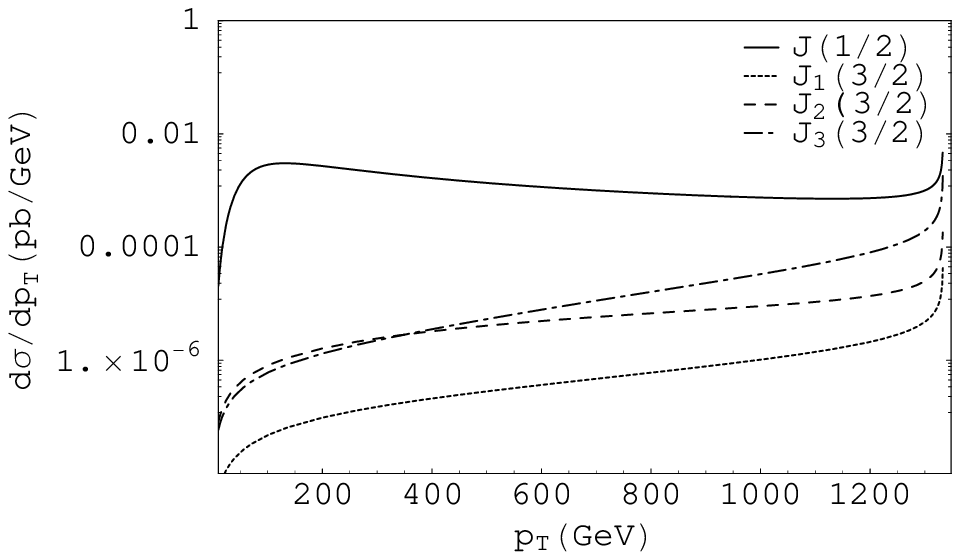}
\caption{The missing $p_{T}$ distribution taking into account
the coupling $c_{iV}^Z=c_{iA}^Z=0.5$ for spin-3/2 currents and $f=f'=1$ for spin-1/2 currents
with $\Lambda=m^*$ where we take $m^*=350$ GeV for ILC and
$m^*=1$ TeV for CLIC.}
\label{fig8}
\end{figure}

\begin{figure}[tbph]
\includegraphics[
height=6cm,
width=8cm
]{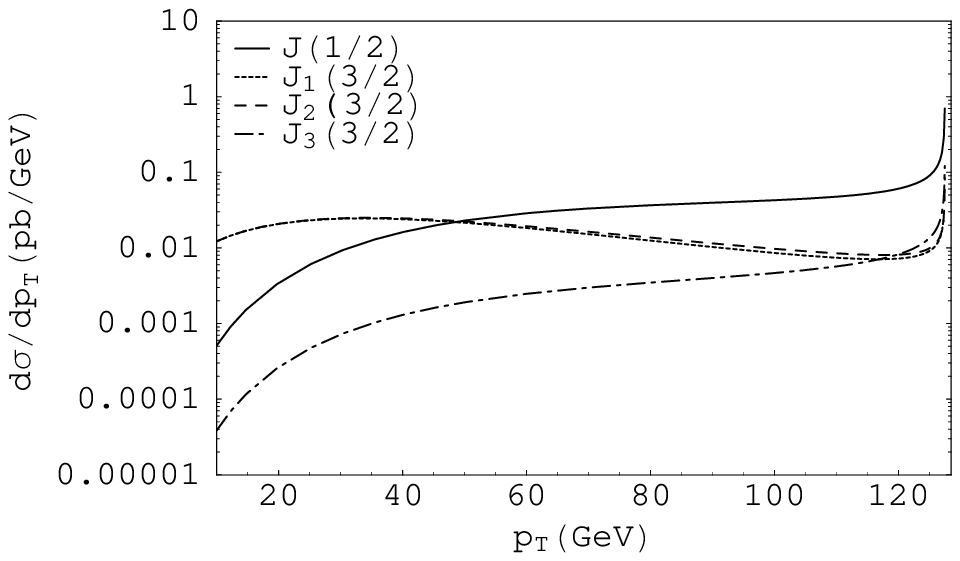}\includegraphics[
height=6cm,
width=8cm
]{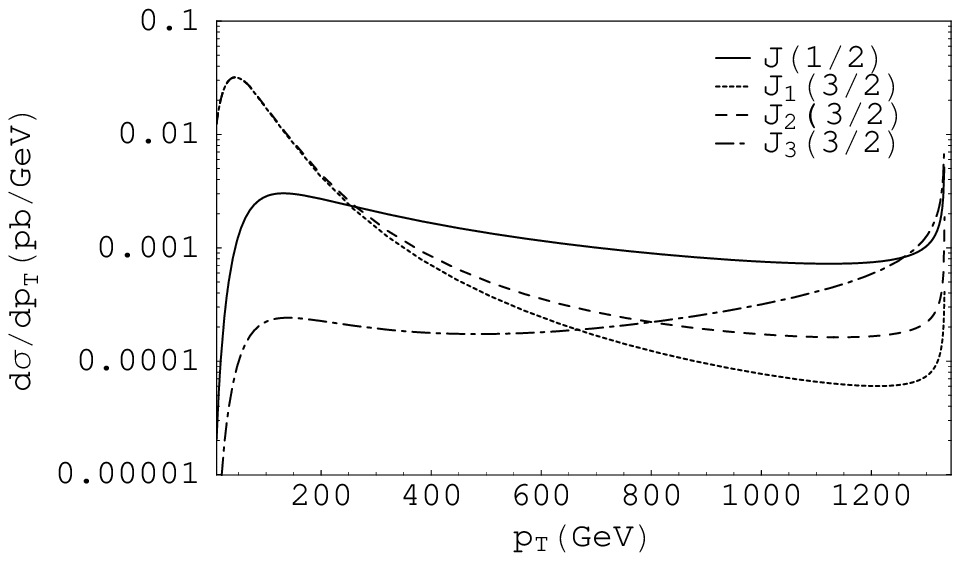}
\caption{The missing $p_{T}$ distribution taking into account
the coupling $c_{iV}^W=c_{iA}^W=0.5$ for the currents $J_i(3/2)$ and
$f=-f'=1$ for $J(1/2)$ with $\Lambda=m^*$ where we used $m^*=350$ GeV for ILC and
$m^*=1$ TeV for CLIC.}
\label{fig9}
\end{figure}

\begin{figure}[tbph]
\includegraphics[height=6cm,width=8cm]{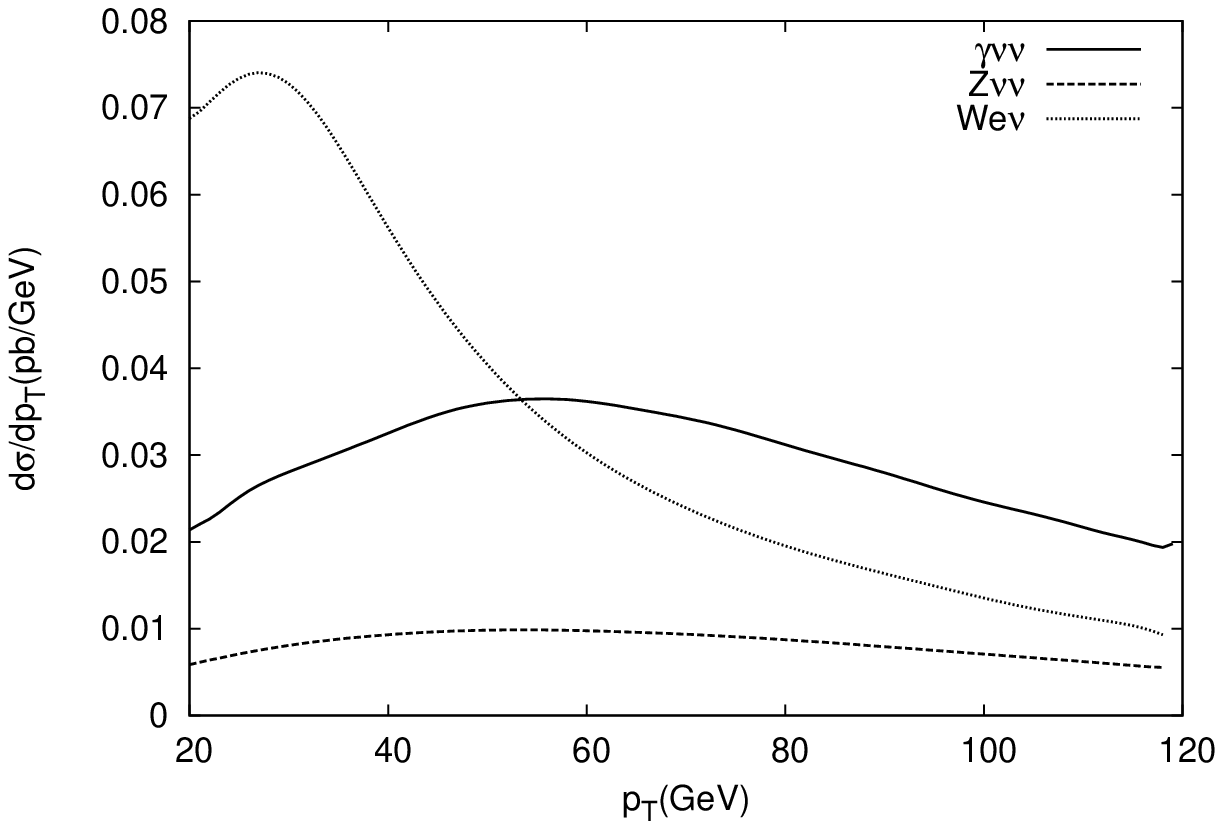}
\includegraphics[height=6cm,width=8cm]{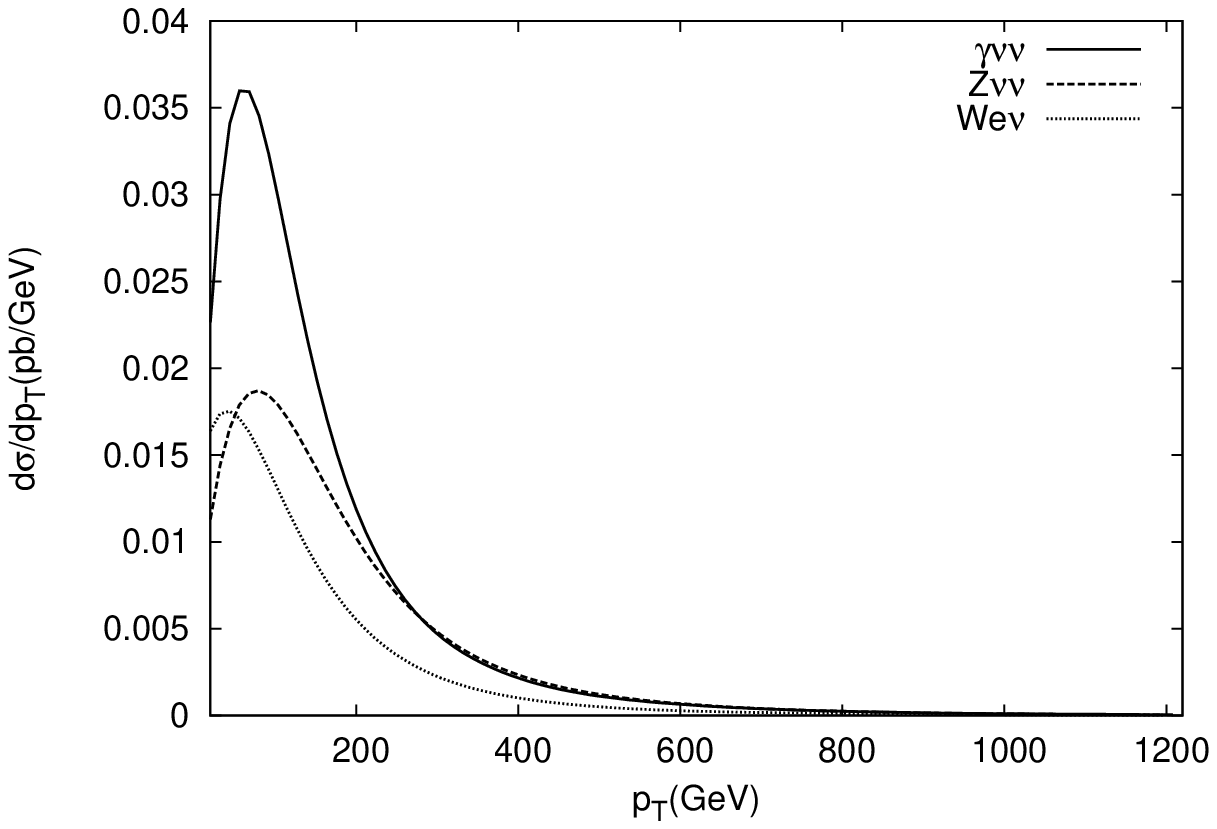}
\caption{Missing $p_T$ distributions of background processes $e^+e^-\to\gamma\nu\bar{\nu}$, $e^+e^-\to Z\nu\bar{\nu}$ and $e^+e^-\to W^+e^-\bar{\nu}$ at ILC (left) and CLIC (right) energies. }
\label{fig10}
\end{figure}

In order to differentiate
the spin-3/2 and spin-1/2 excited neutrino signals we plot the differential
cross sections as a function of $p_T$ in the Figs. \ref{fig7}-\ref{fig9}. The
distributions of missing $p_T$ for the single production of the spin-1/2 and
spin-3/2 excited neutrino show different behaviours. Furthermore, the
production of the excited spin-3/2 neutrino with $J_3$ interactions leads to
different missing $p_T$ distributions than the others ($J_1$ and $J_2$). Concerning
the final state we consider three decay channels of signal $\nu^{\ast
}\rightarrow\nu\gamma$, $\nu^{\ast
}\rightarrow e^{-}W^{+}$ and $\nu^{\ast}\rightarrow\nu Z$  for excited neutrino.
These lead to the final state with $\gamma+\not\!p_T$, $e^\pm+2j+\not\!p_T$ and
$\ell^+\ell^-+\not\!p_T$  as shown in Table~\ref{table2}. At the ILC energy of $\sqrt{s}=500$ GeV,
the shape of the missing transverse momentum $\not\!p_T$ for the signal concerning
$e^+e^-\to W^+e^-\bar{\nu}$ process behaves different from
that of the corresponding background in
Fig. \ref{fig10}. However,
at higher center of mass energies the currents $J_1(3/2)$
and $J_2(3/2)$ show similar behaviour with the
background especially at low $\not\!p_T$ region.

\section{Analysis}

In order to perceive the excited
neutrino signals from the background we can put some cuts on the final state
observable particles, and a cut on the missing transverse momentum. In
general, by applying suitable cuts the signal can be more pronounced over
the background. For the acceptance of interested events we apply the
following initial cuts:
\begin{align}
p_{T}^{e,\gamma }& >20\text{ GeV} \\
\left\vert \eta _{e,\gamma }\right\vert & <2.5
\end{align}
where $p_{T}$ is the transverse momentum of final state detectable particle,
$\eta $ denotes for pseudorapidity. Since the neutrino can be identified as
the missing transverse momentum, we can also apply the same cut on the $\not\!p_{T}$.
After applying these cuts we find the total cross section for the SM
background as $1.062$ $(2.049)$ pb, $0.329$ $(2.083)$ pb and $0.918$
$(0.463) $ pb at $\sqrt{s}=0.5(3)$ TeV for the $e^{+}e^{-}\rightarrow \nu
\overline{\nu }\gamma $ , $e^{+}e^{-}\rightarrow \nu \overline{\nu }Z$ and
$e^{+}e^{-}\rightarrow e^{-}\nu W^{+}$ processes, respectively.
The program CalcHEP \cite{Pukhov} is used to calculate the background
cross sections with suitable cuts. In the analysis, we take into account one coupling $c_{iV},c_{iA}$ is
kept free while the others are assumed to vanish. We present
the signal cross sections in Tables \ref{table3}, \ref{table4}, \ref{table5},
\ref{table6}. In Tables \ref{table7} and \ref{table8} we give the cross sections
for the signal and background in the relevant mass intervals.

We examine the single production of excited spin-3/2 and spin-1/2 neutrinos
in the decay channels $\nu \gamma ,$ $\nu Z$ and $e^{\pm }W^{\mp }$.
Concerning the final states we choose $W$ boson decay hadronically and $Z$
boson decay leptonically,
because of the uncertainty from the neutrinos in the final state. The
corresponding backgrounds can be studied by applying the cuts on the
transverse momentum and pseudorapidities of final state leptons and
jets.
Furthermore, a way of extracting the excited neutrino signal is to impose a
cut $\left\vert m_{ejj}-m^{\ast }\right\vert <25$ GeV on the $ejj$ invariant
mass for the charged weak decay channels of $\nu^{\ast}$. This cut can be relaxed
for higher mass values of $\nu^*$. For $m^*>1.5$ TeV we apply an invariant
mass cut $m_{ejj}>1$ TeV for a clean detection of the signal.

\begin{table}[tbp]
\caption{Final states for single excited neutrino production}
\label{table2}
\begin{tabular}{c|cc}
\hline
Decay mode & Production $e^+e^-\to\bar{\nu}\nu^\ast$ &  \\ \hline
Radiative & $\bar{\nu}\nu\gamma$ &  \\ \hline
Charged-current & $\bar{\nu} e^{-}2j$ &  \\ \hline
Neutral-current & $\bar{\nu}\nu l^+ l^-$ &  \\ \hline
\end{tabular}
\end{table}

\begin{table}[tbp]
\caption{The cross section (in pb) for spin-3/2 signal for
 $c_{iV}^{\protect \gamma }=c_{iA}^{\protect\gamma }=0.5$ and spin-1/2
 signal for $f=-f^{^{\prime
}}=1$ at $\protect\sqrt{s}=0.5$ TeV. Here we take
$\Lambda =m^{\ast }$}
\label{table3}
\begin{tabular}{lllll}
\hline
$m^{\ast }$ (GeV) & $J$ $(1/2)$ & $J_{1}(3/2)$ & $J_{2}(3/2)$ & $J_{3}(3/2)$
 \\ \hline
$200$ & $5.36\times 10^{0}$ & $1.85\times 10^{-1}$ & $6.49\times 10^{-1}$ & $2.60\times 10^{0}$  \\ \hline
$250$ & $3.21\times 10^{0}$ & $1.27\times 10^{-1}$ & $1.76\times 10^{-1}$ & $4.90\times 10^{-1}$  \\ \hline
$300$ & $2.05\times 10^{0}$ & $8.50\times 10^{-2}$ & $4.70\times 10^{-2}$ & $9.90\times 10^{-2}$  \\ \hline
$350$ & $1.43\times 10^{0}$ & $5.10\times 10^{-2}$ & $1.10\times 10^{-2}$ & $1.80\times 10^{-2}$  \\ \hline
$400$ & $9.50\times 10^{-1}$ & $2.50\times 10^{-2}$ & $1.17\times 10^{-3}$ & $2.30\times 10^{-3}$  \\ \hline
$475$ & $3.30\times 10^{-1}$ & $1.80\times 10^{-3}$ & $5.10\times 10^{-6}$ & $5.40\times 10^{-6}$  \\ \hline
\end{tabular}
\end{table}

\begin{table}[tbp]
\caption{The signal cross sections (in pb)
for $c_{iV}^{\protect\gamma }=c_{iA}^{\protect\gamma }=0.5$
and $f=-f^{^{\prime}}=1$ for excited neutrinos
at $\protect\sqrt{s}=3$ TeV. Here we take $\Lambda =m^{\ast }$.}
\label{table4}
\begin{tabular}{lllll}
\hline
$m^{\ast \text{ }}$(GeV) & $J(1/2)$ & $J_{1}(3/2)$ & $J_{2}(3/2)$ &
$J_{3}(3/2)$ \\ \hline
$250$ & $4.97\times 10^{0}$ & $6.90\times 10^{-2}$ & $1.73\times 10^{1}$ & $1.35\times 10^{3}$  \\ \hline
$500$ & $1.13\times 10^{0}$ & $2.00\times 10^{-2}$ & $1.00\times 10^{0}$ & $2.01\times 10^{1}$  \\ \hline
$1000$ & $2.80\times 10^{-1}$ & $6.70\times 10^{-3}$ & $4.60\times 10^{-2}$ & $2.50\times 10^{-1}$  \\ \hline
$1500$ & $1.29\times 10^{-1}$ & $3.50\times 10^{-3}$ & $4.90\times 10^{-3}$ & $1.40\times 10^{-2}$  \\ \hline
$2000$ & $7.60\times 10^{-2}$ & $1.70\times 10^{-3}$ & $5.10\times 10^{-4}$ & $9.10\times 10^{-4}$  \\ \hline
$2500$ & $5.00\times 10^{-2}$ & $5.00\times 10^{-4}$ & $2.10\times 10^{-5}$ & $2.70\times 10^{-5}$  \\ \hline
$2750$ & $4.20\times 10^{-2}$ & $1.30\times 10^{-4}$ & $1.14\times 10^{-6}$ & $1.29\times 10^{-6}$  \\ \hline
\end{tabular}
\end{table}

\begin{table}[tbp]
\caption{The
signal cross section (in pb)  for
$c_{iV}^{Z}=c_{iA}^{Z}=0.5$ and $f=f^{^{\prime }}=1$ for the
 excited neutrinos at $\protect\sqrt{s}=0.5$ TeV. Here we take $\Lambda =m^{\ast }$.}
\label{table5}
\begin{tabular}{lllll}
\hline
$m^{\ast }$ (GeV) & $J$ $(1/2)$ & $J_{1}(3/2)$ & $J_{2}(3/2)$ & $J_{3}(3/2)$
 \\ \hline
$200$ & $5.68\times 10^{0}$ & $7.00\times 10^{-2}$ & $2.40\times 10^{-1}$ & $9.90\times 10^{-1}$  \\ \hline
$250$ & $3.82\times 10^{0}$ & $4.80\times 10^{-2}$ & $6.70\times 10^{-2}$ & $1.90\times 10^{-1}$  \\ \hline
$300$ & $2.61\times 10^{0}$ & $3.20\times 10^{-2}$ & $1.80\times 10^{-2}$ & $3.80\times 10^{-2}$  \\ \hline
$350$ & $1.84\times 10^{0}$ & $2.00\times 10^{-2}$ & $4.20\times 10^{-3}$ & $6.90\times 10^{-3}$  \\ \hline
$400$ & $1.27\times 10^{0}$ & $9.50\times 10^{-3}$ & $6.40\times 10^{-4}$ & $8.80\times 10^{-4}$  \\ \hline
$475$ & $4.40\times 10^{-1}$ & $6.80\times 10^{-4}$ & $1.92\times 10^{-6}$ & $2.07\times 10^{-6}$  \\ \hline
\end{tabular}
\end{table}

\begin{table}[tbp]
\caption{The cross sections (in pb)
for $c_{iV}^{Z}=c_{iA}^{Z}=0.5$
and $f=f^{^{\prime }}=1$ for excited neutrinos at
$\protect\sqrt{s}=3$ TeV, and $\Lambda =m^{\ast }$}
\label{table6}
\begin{tabular}{lllll}
\hline
$m^{\ast \text{ }}$(GeV) & $J(1/2)$ & $J_{1}(3/2)$ & $J_{2}(3/2)$ & $%
J_{3}(3/2)$  \\ \hline
$250$ & $5.95\times 10^{0}$ & $2.50\times 10^{-2}$ & $6.15\times 10^{0}$ & $4.80\times 10^{2}$  \\ \hline
$500$ & $1.50\times 10^{0}$ & $7.00\times 10^{-3}$ & $3.60\times 10^{-1}$ & $7.16\times 10^{0}$  \\ \hline
$1000$ & $3.90\times 10^{-1}$ & $2.40\times 10^{-3}$ & $1.60\times 10^{-2}$ & $9.00\times 10^{-2}$  \\ \hline
$1500$ & $1.79\times 10^{-1}$ & $1.30\times 10^{-3}$ & $1.70\times 10^{-3}$ & $4.80\times 10^{-3}$  \\ \hline
$2000$ & $1.05\times 10^{-1}$ & $6.10\times 10^{-4}$ & $1.80\times 10^{-4}$ & $3.20\times 10^{-4}$  \\ \hline
$2500$ & $7.00\times 10^{-2}$ & $1.80\times 10^{-4}$ & $7.54\times 10^{-6}$ & $9.74\times 10^{-6}$  \\ \hline
$2750$ & $5.90\times 10^{-2}$ & $4.80\times 10^{-5}$ & $4.10\times 10^{-7}$ & $4.60\times 10^{-7}$  \\ \hline
\end{tabular}
\end{table}

\begin{table}[tbp]
\caption{The signal and background cross sections (in pb) after the cuts. The
signal cross section is given for $\Lambda =m^{\ast }$ and
$c_{iV}^{W}=c_{iA}^{W}=0.5$ for the spin-3/2 and $f=-f^{^{\prime }}=1$ for
the spin-1/2 at $\protect\sqrt{s}=0.5$ TeV. We have used $\Delta m=50$ GeV.}
\label{table7}
\begin{tabular}{llllll}
\hline
$m^{\ast }$ (GeV) & $J$ $(1/2)$ & $J_{1}(3/2)$ & $J_{2}(3/2)$ & $J_{3}(3/2)$
& $\sigma_{B}$($\Delta m$)\\ \hline
$200$ & $8.82\times 10^{0}$ & $2.30\times 10^{0}$ & $3.61\times 10^{0}$ & $7.70\times 10^{0}$ & $9.36\times 10^{-2}$ \\ \hline
$250$ & $5.81\times 10^{0}$ & $1.83\times 10^{0}$ & $2.39\times 10^{0}$ & $2.75\times 10^{0}$ & $9.12\times 10^{-2}$ \\ \hline
$300$ & $4.02\times 10^{0}$ & $1.52\times 10^{0}$ & $1.76\times 10^{0}$ & $1.25\times 10^{0}$ & $9.65\times 10^{-2}$ \\ \hline
$350$ & $2.82\times 10^{0}$ & $1.26\times 10^{0}$ & $1.32\times 10^{0}$ & $6.60\times 10^{-1}$ & $1.21\times 10^{-1}$ \\ \hline
$400$ & $1.97\times 10^{0}$ & $9.50\times 10^{-1}$ & $9.10\times 10^{-1}$ & $3.60\times 10^{-1}$ & $1.69\times 10^{-1}$ \\ \hline
$475$ & $6.70\times 10^{-1}$ & $2.80\times 10^{-1}$ & $2.10\times 10^{-1}$ & $9.50\times 10^{-2}$ & $1.23\times 10^{-1}$ \\ \hline
\end{tabular}
\end{table}

\begin{table}[tbp]
\caption{The signal and background cross sections (in pb) after the cuts. The
cross section is given for $\Lambda =m^{\ast }$and
$c_{iV}^{W}=c_{iA}^{W}=0.5$ for the spin-3/2 and $f=-f^{^{\prime }}=1$ for
the spin-1/2 signal at $\protect\sqrt{s}=3$ TeV.}
\label{table8}
\begin{tabular}{llllll}
\hline
$m^{\ast \text{ }}$(GeV) & $J(1/2)$ & $J_{1}(3/2)$ & $J_{2}(3/2)$ &
$J_{3}(3/2)$ & $\sigma_B$($\Delta m$)\\ \hline
$250$ & $9.01\times 10^{0}$ & $1.23\times 10^{0}$ & $2.08\times 10^{1}$ & $1.54\times 10^{3}$ & $1.96\times 10^{-2}$ \\ \hline
$500$ & $2.35\times 10^{0}$ & $3.70\times 10^{-1}$ & $1.70\times 10^{0}$ & $2.69\times 10^{1}$ & $1.69\times 10^{-2}$ \\ \hline
$1000$ & $6.00\times 10^{-1}$ & $1.60\times 10^{-1}$ & $2.70\times 10^{-1}$ & $6.50\times 10^{-1}$ & $1.00\times 10^{-2}$ \\ \hline
$1500$ & $2.76\times 10^{-1}$ & $1.20\times 10^{-1}$ & $1.50\times 10^{-1}$ & $1.00\times 10^{-1}$ & $1.83\times 10^{-1}$ \\ \hline
$2000$ & $1.62\times 10^{-1}$ & $1.10\times 10^{-1}$ & $1.20\times 10^{-1}$ & $3.50\times 10^{-2}$ & $1.83\times 10^{-1}$ \\ \hline
$2500$ & $1.08\times 10^{-1}$ & $1.00\times 10^{-1}$ & $1.00\times 10^{-1}$ & $1.70\times 10^{-2}$ & $1.83\times 10^{-1}$ \\ \hline
$2750$ & $9.00\times 10^{-2}$ & $9.70\times 10^{-2}$ & $9.00\times 10^{-2}$ & $1.30\times 10^{-2}$ & $1.83\times 10^{-1}$ \\ \hline
\end{tabular}
\end{table}

\begin{figure}[tbph]
\includegraphics[height=6cm,width=8cm]
{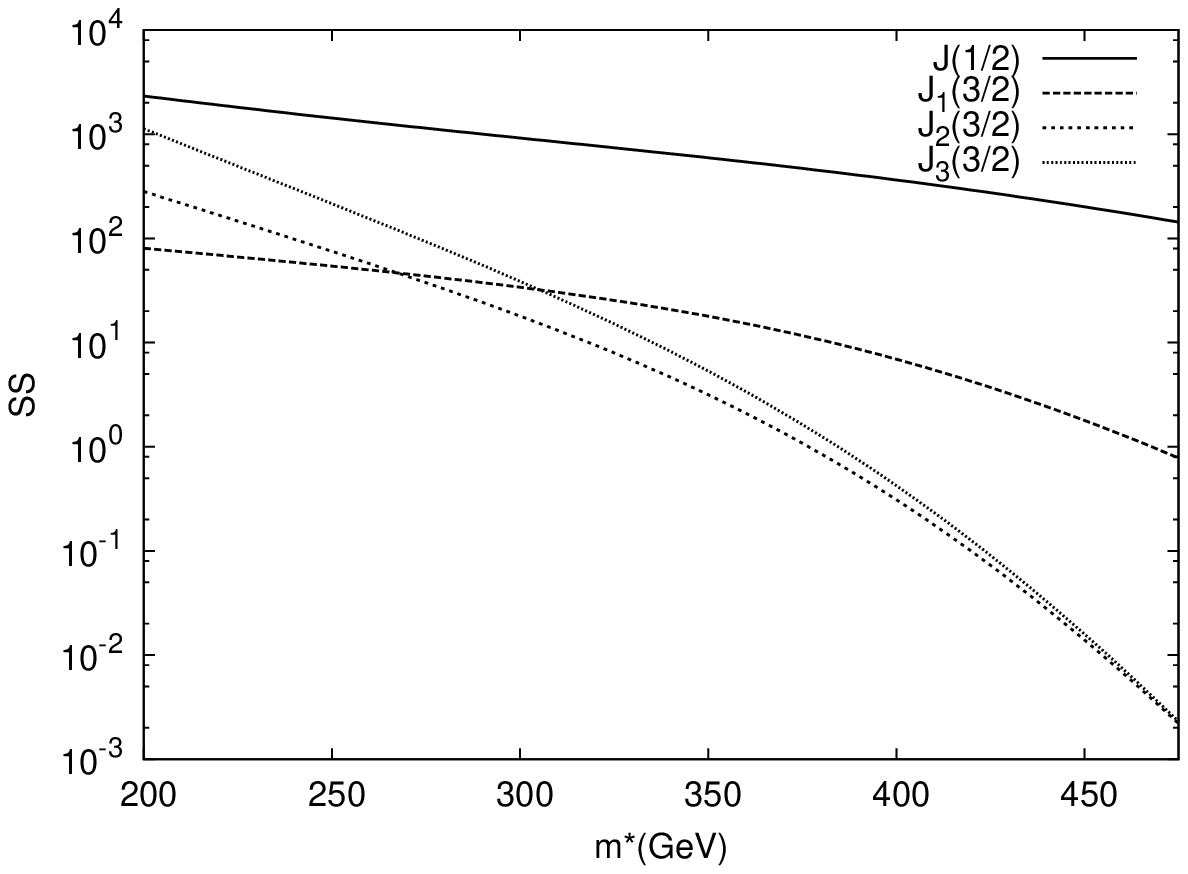} \includegraphics[height=6cm,width=8cm]
{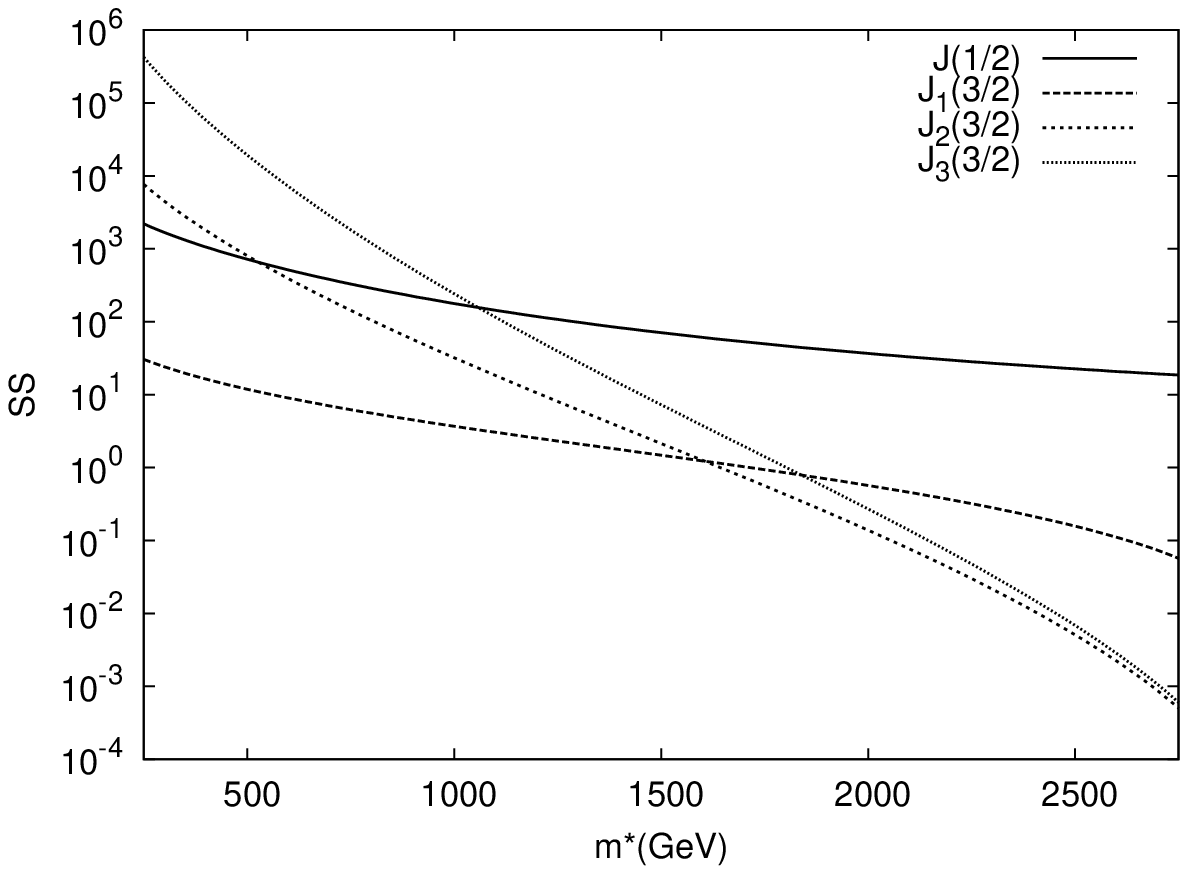}
\caption{Statistical significance as a function of excited neutrino mass for
different spin-3/2 currents with the couplings $c_{iV}^{\protect\gamma %
}=c_{iA}^{\protect\gamma }=0.5,$ and spin-1/2 current with $f=-f^{^{\prime
}}=1$ at $\protect\sqrt{s}=0.5$ TeV(first panel) and $\protect\sqrt{s}=3$
TeV (second panel) using an integrated luminosity of $L_{int}=200$ fb$^{-1}$
for ILC and $L_{int}=400$ fb$^{-1}$ for CLIC\ energies.}
\label{fig11}
\end{figure}

\begin{figure}[tbph]
\includegraphics[height=6cm,width=8cm]
{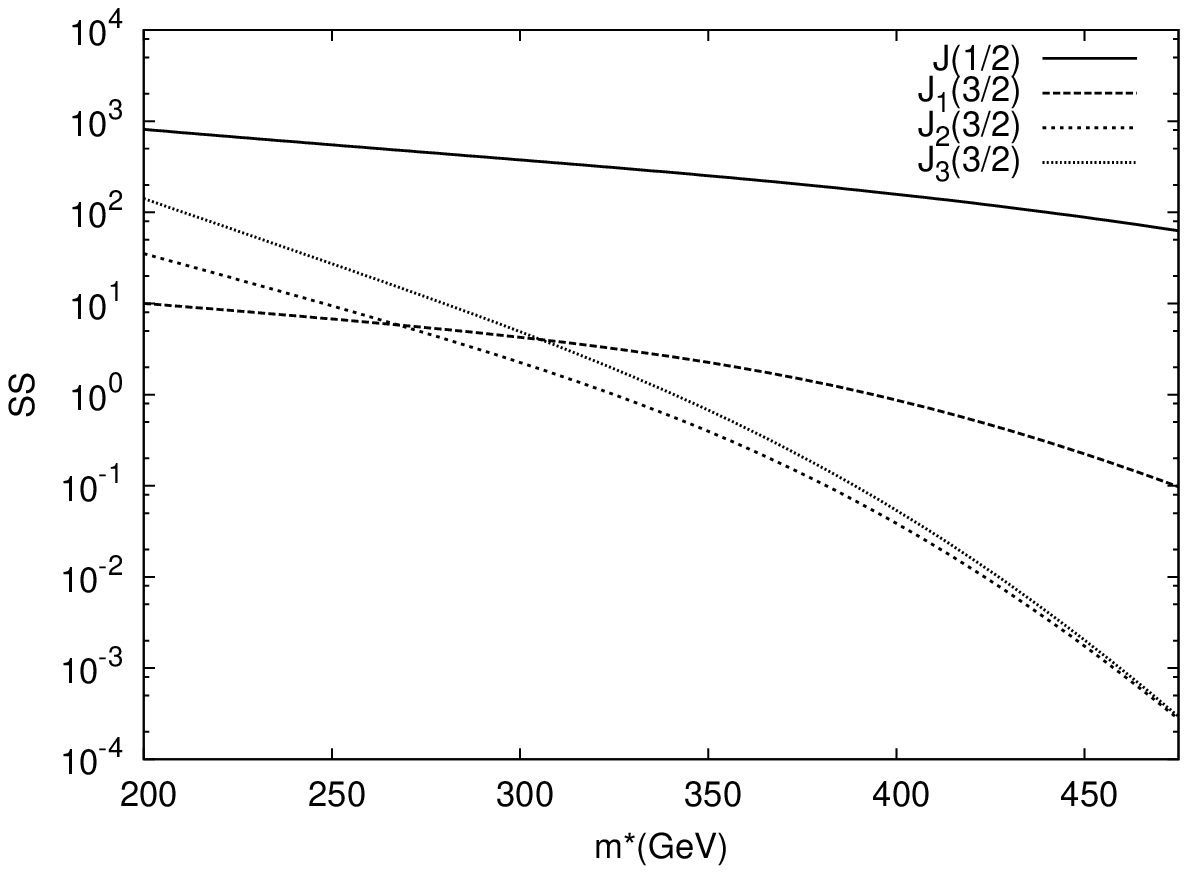} \includegraphics[height=6cm,width=8cm]
{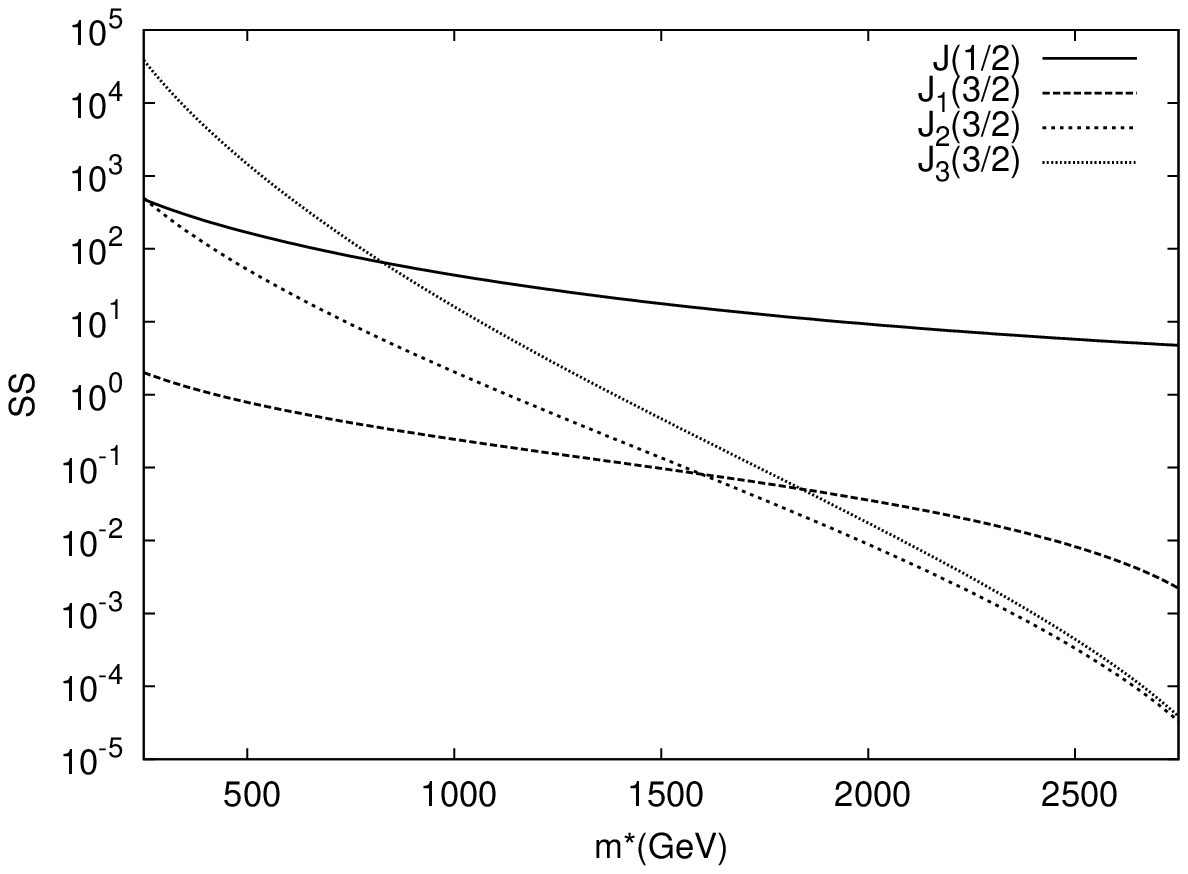}
\caption{Statistical significance as a function of excited neutrino mass for
different spin-3/2 currents with the couplings $c_{iV}^{Z}=c_{iA}^{Z}=0.5,$%
and spin-1/2 current with $f=f^{^{\prime }}=1$ at $\protect\sqrt{s}=0.5$
TeV(first panel) and $\protect\sqrt{s}=3$ TeV (second panel) using an
integrated luminosity of $L_{int}=200$ fb$^{-1}$ for ILC and $L_{int}=400$ fb%
$^{-1}$ for CLIC\ energies.}
\label{fig12}
\end{figure}

\begin{figure}[tbph]
\includegraphics[height=6cm,width=8cm]
{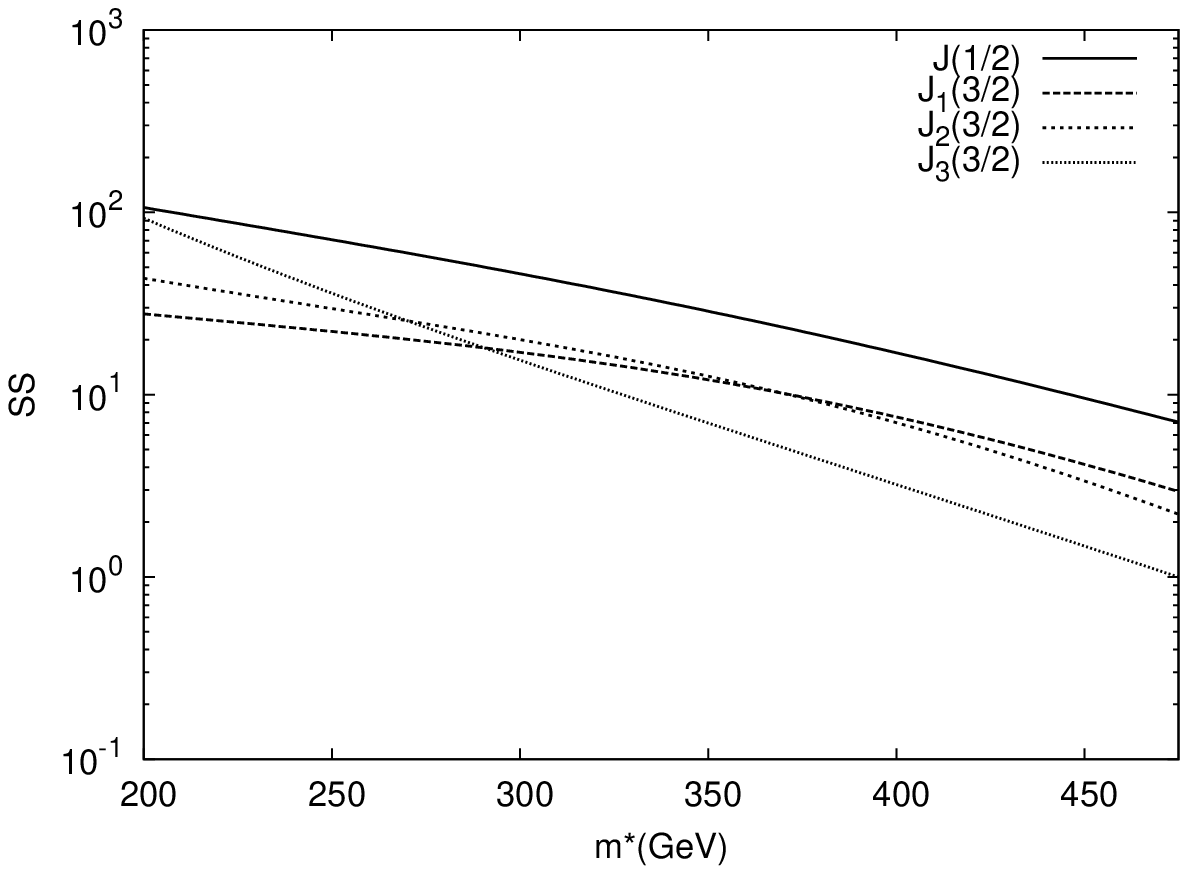} \includegraphics[height=6cm,width=8cm]
{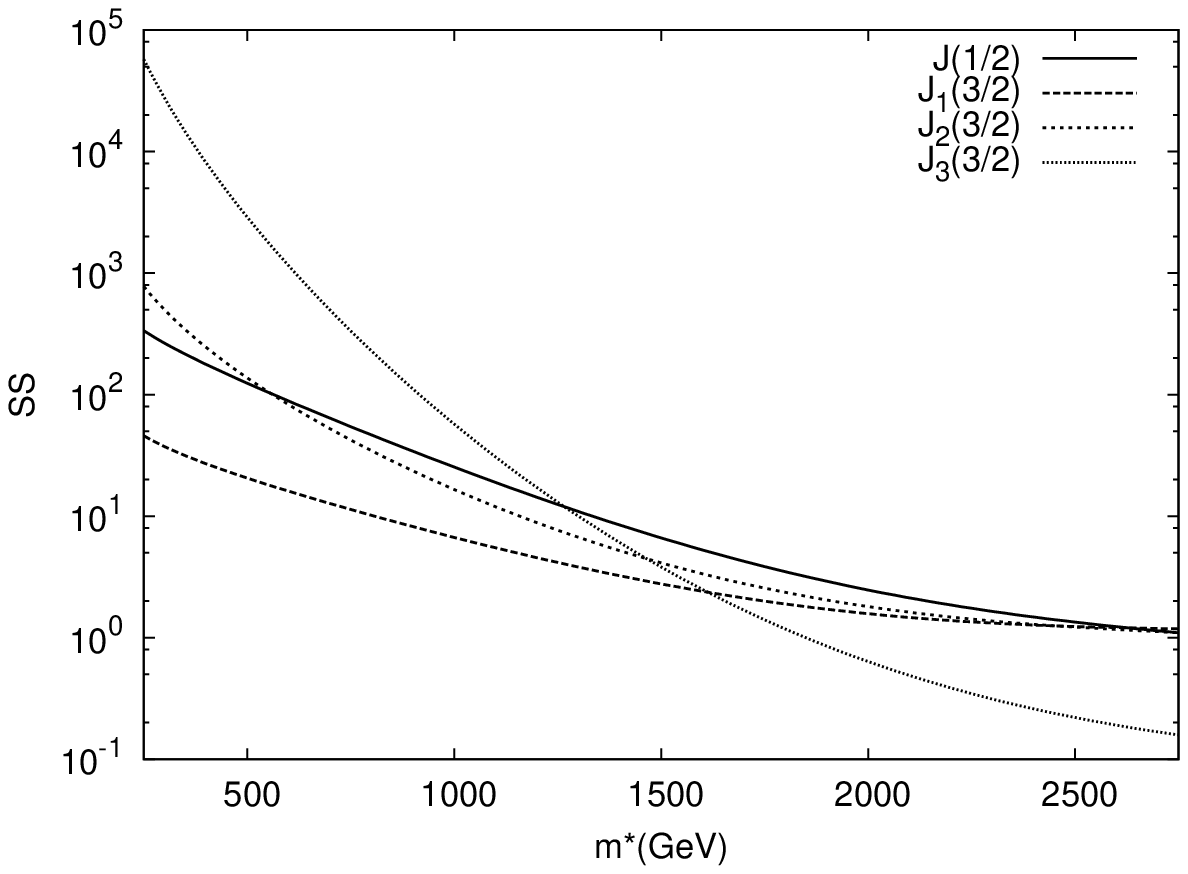}
\caption{Statistical significance as a function of excited neutrino mass for
different spin-3/2 currents with the couplings $c_{iV}^{W}=c_{iA}^{W}=0.05,$
and spin-1/2 current with $f=-f^{^{\prime }}=1$ at $\protect\sqrt{s}=0.5$
TeV(first panel) and $\protect\sqrt{s}=3$ TeV (second panel) using an
integrated luminosity of $L_{int}=200$ fb$^{-1}$ for ILC and $L_{int}=400$ fb
$^{-1}$ for CLIC\ energies.}
\label{fig13}
\end{figure}

For the analysis we
define the statistical significance ($SS$) of the signal as
\begin{equation*}
SS=\frac{\sigma_{S}}{\sqrt{\sigma_{B}}}\sqrt{\epsilon.L}
\end{equation*}
where $L_{int}$ is the integrated luminosity of the collider and $\epsilon$
is the efficiency to detect the signal in the chosen channel.
In the $\nu^*\to \nu\gamma$ and $\nu^*\to\nu Z$ channel,
we have used the cross section for the signal and background to
calculate the significance. Requiring $SS>3$ the attainable mass range covers
up to $m^*\approx 1.3-1.5$ TeV
(as seen from Fig.~\ref{fig11}-\ref{fig13} )
for an excited spin-3/2 neutrino at CLIC with $\sqrt{s}=3$ TeV.
For the $\nu^*\to eW$ channel, the values of
the significance are evaluted at each mass values
(invariant mass intervals) and the results
are presented in Fig. \ref{fig13} the discovery reach for excited spin-3/2
and spin-1/2 neutrinos.
A smaller coupling $c_{iV}^W=c_{iA}^W=0.05$ can be
reached in the $\nu^*\to eW$ channel. At the ILC, the attainable mass ranges for spin-3/2 charged currents can be covered up to the center of mass energy.

\section{Conclusion}

We consider only the gauge interactions of excited spin-1/2 and spin-3/2
particles with the SM particles. Our analysis show that spin-3/2 excited
neutrinos can be easily separated from the spin-1/2 ones by examining
both the transverse momentum distributions of final state
visible and invisible particles when they produced singly at future linear
colliders. If polarized $e^{+}$ and $e^{-}$ beams are used the chiral structure
of the couplings can be identified and more precise measurements can be
performed at the ILC and CLIC environment.
A previous work on excited spin-3/2 electrons \cite{Cakir08}
shows that they can be distinguished from the spin-1/2 ones by analyzing the
angular distribution of final state observable particles.

Excited neutrinos can come in three families, $\nu _{e}^{\ast },$ $\nu _{\mu
}^{\ast }$ and $\nu _{\tau }^{\ast }$, this study for $\nu _{e}^{\ast }$ can also
be enlarged by applying similar analysis to the excited muon-neutrinos and
tau-neutrinos at the future high energy linear colliders, and their signal
distribution would be very different from that for the $\nu_{e^*}$ when
they are produced in the $t$-channel diagrams with the LFV.

\subsection{Appendix}
\appendix

In equation (12) the operator terms $T_{ij}^{(1)}$ and propogators $%
P_{ij}^{(1)}$ for the spin-3/2 excited neutrino interaction current $J_{1}$
are given as

\begin{center}
\begin{align*}
T_{11}^{(1)} & =-{g}_{{e}}^{2\,}({c}_{1A}^{\gamma}{}^{2}+{c}_{1V}^{\gamma
}{}^{2})({m}^{\ast2}-s)\,\left( -t\,\left( s+t\right) +{m}^{\ast 2}\,\left(
2\,s+t\right) \right) \\
T_{12}^{(1)} & =[-{g}_{{e}}^{\,}\,{g}_{{z}}(m_{Z}^{2}-s)({c}_{1A}^{\gamma}({c%
}_{1V}^{Z}{c}_{A}^{f}{m}^{\ast2}s({m}^{\ast2}-s-2t)+{c}_{1A}^{Z}{c}%
_{V}^{f}(st(s+t)+{m}^{\ast4}(2s+t) \\
& -{m}^{\ast2}(2s^{2}+2st+t^{2})))+{c}_{1V}^{\gamma}({c}_{1A}^{Z}{c}_{A}^{f}{%
m}^{\ast2}s({m}^{\ast2}-s-2t) \\
& +{c}_{1V}^{Z}{c}_{V}^{f}(st(s+t)+{m}^{\ast4}(2s+t)-{m}^{%
\ast2}(2s^{2}+2st+t^{2}))))/2 \\
T_{13}^{(1)} & =[-{g}_{{e}}^{\,}{g}_{{w}}^{\,}({c}_{1A}^{W}{}-{c}%
_{1V}^{W}{})({c}_{1A}^{\gamma}{}-{c}_{1V}^{\gamma}{})(({m}^{\ast 4}+st)(s+t)-%
{m}^{\ast2}(s^{2}+t^{2}))]/4\sqrt{2} \\
T_{22}^{(1)} & =[-{g}_{{z}}^{2}{{(4}c}_{1A}^{Z}\,{c}_{1V}^{Z}\,{c}_{A}^{f}{c}%
_{V}^{f}\,{m}^{\ast2}s({m}^{\ast2}-s-2t) \\
& +({c}_{1A}^{Z}{}^{2}+{c}_{1V}^{Z}{}^{2})(({c}_{A}^{f2}+{c}_{V}^{f2})\left(
{m}^{\ast2}-s\right) (-t(s+t)+{m}^{\ast2}(2s+t)))]/4 \\
T_{23}^{(1)} & =[-{g}_{{e}}^{\,}\,{g}_{{w}}(({c}_{1A}^{W}{}-{c}_{1V}^{W}{})({%
c}_{1A}^{Z}{}-{c}_{1V}^{Z}{})({c}_{A}^{f}+{c}_{V}^{f})\left( {m}_{{z}%
}^{2}-s\right) \\
& (({m}^{\ast4}+st)(s+t)-{m}^{\ast2}(s^{2}+t^{2})))]/8\sqrt{2} \\
T_{33}^{(1)} & =[-{g}_{{w}}^{2\,}(({c}_{1A}^{W}{}^{2}+{c}_{1V}^{W}{}^{2})(-{m%
}^{\ast2}+t)(s(s+t)-{m}^{\ast2}(s+2t))) \\
& +2{c}_{1A}^{W}{}{c}_{1V}^{W}{}{m}^{\ast2}({m}^{\ast2}-2s-t)t]/4 \\
& \text{and} \\
P_{11}^{(1)} & =1/s^{2},P_{12}^{(1)}=1/s({m}_{{z}}^{4}+s^{2}+{m}_{{z}%
}^{2}(-2s+\Gamma_{Z}^{2})),\text{ }P_{13}^{(1)}=1/s({m}_{{w}}^{2}-t), \\
P_{22}^{(1)} & =1/({m}_{{z}}^{4}+s^{2}+{m}_{{z}}^{2}(-2s+\Gamma _{Z}^{2})),
\\
P_{23}^{(1)} & =1/(({m}_{{w}}^{2}-t)({m}_{{z}}^{4}+s^{2}+{m}_{{z}%
}^{2}(-2s+\Gamma_{Z}^{2}))),\text{ }P_{33}^{(1)}=1/({m}_{{w}}^{2}-t)^{2}
\end{align*}

The $T_{ij}^{(2)}$ and $P_{ij}^{(2)}$ terms for the current $J_{2}$ are
given by

\begin{align*}
T_{11}^{(2)} & =-{g}_{{e}}^{2\,}({c}_{2A}^{\gamma}{}^{2}+{c}_{2V}^{\gamma
}{}^{2})({m}^{\ast2}-s)^{2}\,\left( -s^{2}-2st\,-2t^{2}+{m}^{\ast 2}\,\left(
\,s+2t\right) \right) /2 \\
T_{12}^{(2)} & =[{g}_{{e}}^{\,}\,{g}_{{z}}(m^{\ast2}-s)^{2}\left( -{m}_{{z}%
}^{2}+s\right) ({c}_{2A}^{\gamma}({c}_{2V}^{Z}{c}_{A}^{f}s({m}^{\ast2}-s-2t)
\\
& +{c}_{2A}^{Z}{c}_{V}^{f}(-s^{2}-2st-2t^{2}+{m}^{\ast2}(s+2t))) \\
& +{c}_{2V}^{\gamma}({c}_{2A}^{Z}{c}_{A}^{f}s({m}^{\ast2}-s-2t)+{c}%
_{2V}^{Z}{}{c}_{V}^{f}(-s^{2}-2st-2t^{2}+{m}^{\ast2}(s+2t))))]/4 \\
T_{13}^{(2)} & =[-{g}_{{e}}^{\,}{g}_{{w}}^{\,}({c}_{2A}^{W}{}+{c}%
_{2V}^{W}{})({c}_{2A}^{\gamma}{}+{c}_{2V}^{\gamma}{})({m}^{\ast2}-s-t)(s+t)({%
m}^{\ast4}-st-{m}^{\ast2}(s+t))]/4\sqrt{2} \\
T_{22}^{(2)} & =[-{g}_{{z}}^{2}({m}^{\ast2}-s)^{2}({{-4}c}_{2A}^{Z}\,{c}%
_{2V}^{Z}\,{c}_{A}^{f}{c}_{V}^{f}s\,(-{m}^{\ast2}+s+2t) \\
& +({c}_{2A}^{Z}{}^{2}+{c}_{2V}^{Z}{}^{2})(({c}_{A}^{f2}+{c}%
_{V}^{f2})(-s^{2}-2st-2t^{2}+{m}^{\ast2}(s+2t))))]/8 \\
T_{23}^{(2)} & =[-{g}_{{z}}^{\,}\,{g}_{{w}}(({c}_{2A}^{W}{}+{c}_{2V}^{W}{})({%
c}_{2A}^{Z}{}+{c}_{2V}^{Z}{})({c}_{A}^{f}+{c}_{V}^{f})\left( {m}_{{z}%
}^{2}-s\right) (s+t) \\
& (-{m}^{\ast2}+s+t)(s+t)(-{m}^{\ast4}+st+{m}^{\ast2}(s+t)))]/8\sqrt {2} \\
T_{33}^{(2)} & =[-{g}_{{w}}^{2\,}((-{m}^{\ast2}+t)^{2}(2{c}_{2A}^{W}{}{c}%
_{2V}^{W}{}({m}^{\ast2}-2s-t)t \\
& +({c}_{2A}^{W}{}^{2}+{c}_{2V}^{W}{}^{2})(-2s^{2}-2st-t^{2}+{m}^{\ast
2}(2s+t))))]/4 \\
& \text{and} \\
P_{11}^{(2)} & =1/s^{2}\Lambda^{2},P_{12}^{(2)}=1/s({m}_{{z}}^{4}+s^{2}+{m}_{%
{z}}^{2}(-2s+\Gamma_{Z}^{2}))\Lambda^{2},\text{ }P_{13}^{(2)}=1/s({m}_{{w}%
}^{2}-t)\Lambda^{2}, \\
P_{22}^{(2)} & =1/({m}_{{z}}^{4}+s^{2}+{m}_{{z}}^{2}(-2s+\Gamma
_{Z}^{2}))\Lambda^{2}, \\
P_{23}^{(2)} & =1/(({m}_{{w}}^{2}-t)({m}_{{z}}^{4}+s^{2}+{m}_{{z}%
}^{2}(-2s+\Gamma_{Z}^{2})))\Lambda^{2},\text{ }P_{33}^{(2)}=1/({m}_{{w}%
}^{2}-t)^{2}\Lambda^{2}
\end{align*}

The $T_{ij}^{(3)}$ and $P_{ij}^{(3)}$ terms for the current $J_{3}$ are
given by

\begin{align*}
T_{11}^{(3)} & =-{g}_{{e}}^{2\,}({c}_{3A}^{\gamma}{}^{2}+{c}_{3V}^{\gamma
}{}^{2})({m}^{\ast2}-s)^{2}\,\left( {m}^{\ast4}+2t(s+t)-{m}^{\ast 2}\,\left(
\,s+2t\right) \right) /2 \\
T_{12}^{(3)} & =[-{g}_{{e}}^{\,}\,{g}_{{z}}(m^{\ast2}-s)^{2}\left( -{m}_{{z}%
}^{2}+s\right) ({c}_{3A}^{\gamma}({c}_{3V}^{Z}{c}_{A}^{f}{m}^{\ast2}({m}%
^{\ast2}-s-2t)- \\
& {c}_{3A}^{Z}{c}_{V}^{f}({m}^{\ast4}+2t(s+t)-{m}^{\ast2}(s+2t)))+{c}%
_{3V}^{\gamma}({c}_{3A}^{Z}{c}_{A}^{f}{m}^{\ast2}({m}^{\ast2}-s-2t)- \\
& {c}_{3V}^{Z}{c}_{V}^{f}({m}^{\ast4}+2t(s+t)-{m}^{\ast2}(s+2t))))]/4 \\
T_{13}^{(3)} & =[-{g}_{{e}}^{\,}{g}_{{w}}^{\,}({c}_{3A}^{W}{}-{c}%
_{3V}^{W}{})({c}_{3A}^{\gamma}{}-{c}_{3V}^{\gamma}{})({m}^{\ast2}-s-t)t({m}%
^{\ast2}(s-t)+st+{m}^{\ast2}(-s+t))]/4\sqrt{2} \\
T_{22}^{(3)} & =[-{g}_{{z}}^{2}({m}^{\ast2}-s)^{2}s({{4}c}_{3A}^{Z}\,{c}%
_{3V}^{Z}\,{c}_{A}^{f}{c}_{V}^{f}{m}^{\ast2}\,(-{m}^{\ast2}+s+2t) \\
& +({c}_{3A}^{Z}{}^{2}+{c}_{3V}^{Z}{}^{2})(({c}_{A}^{f2}+{c}_{V}^{f2})({m}%
^{\ast4}+2t(s+t)-{m}^{\ast2}(s+2t)))]/8 \\
T_{23}^{(3)} & =[{g}_{{z}}^{\,}\,{g}_{{w}}s(({c}_{3A}^{W}{}-{c}_{3V}^{W}{})({%
c}_{3A}^{Z}{}-{c}_{3V}^{Z}{})({c}_{A}^{f}+{c}_{V}^{f})\left( {m}_{{z}%
}^{2}-s\right) t \\
& (-{m}^{\ast2}+s+t)(-{m}^{\ast2}(s-t)+st+{m}^{\ast2}(-s+t)))]/8\sqrt {2} \\
T_{33}^{(3)} & =[-{g}_{{w}}^{2\,}((-{m}^{\ast2}+t)^{2}t(2{c}_{3A}^{W}{}{c}%
_{3V}^{W}{}{m}^{\ast2}(-{m}^{\ast2}+2s+t) \\
& +(({c}_{3A}^{W}{}^{2}+{c}_{3V}^{W}{}^{2})({m}^{\ast4}+2s(s+t)-{m}%
^{\ast2}(2s+t))))]/4 \\
& \text{and} \\
P_{11}^{(3)} & =1/s\Lambda^{4},P_{12}^{(3)}=1/({m}_{{z}}^{4}+s^{2}+{m}_{{z}%
}^{2}(-2s+\Gamma_{Z}^{2}))\Lambda^{4},\text{ }P_{13}^{(3)}=1/(t-{m}_{{w}%
}^{2})\Lambda^{4}, \\
P_{22}^{(3)} & =1/({m}_{{z}}^{4}+s^{2}+{m}_{{z}}^{2}(-2s+\Gamma
_{Z}^{2}))\Lambda^{4}, \\
P_{23}^{(3)} & =1/((t-{m}_{{w}}^{2})({m}_{{z}}^{4}+s^{2}+{m}_{{z}%
}^{2}(-2s+\Gamma_{Z}^{2})))\Lambda^{4},\text{ }P_{33}^{(3)}=1/(t-{m}_{{w}%
}^{2})^{2}\Lambda^{4}
\end{align*}
\end{center}

\begin{acknowledgments}
We thank A. Belyaev and A. Pukhov for discussions on the CalcHEP implementation. 
OC acknowledges the support from the CERN TH Division.
AO would like to thank to the members of the Nuclear Theory Group
of University of Wisconsin for their hospitality. AO acknowledges
the support through the Scientific and Technical Research Council
(TUBITAK) BIDEB-2214.
This work was partially supported by the Turkish State Planning
Organisation (DPT) under grant No DPT-2006K-120470.
\end{acknowledgments}

\end{document}